\documentclass[aps,prd,twocolumn,superscriptaddress,showpacs,amsmath,amssymb,amsthm,nofootinbib]{revtex4-2}
\usepackage{amsmath,amssymb}
\usepackage{enumitem}  
\usepackage[usenames, dvipsnames]{color} 
\usepackage{graphicx}     
\usepackage{subcaption}
\usepackage[normalem]{ulem}
\usepackage[utf8]{inputenc}
\usepackage{url}
 \usepackage{xcolor}

\newcommand{\nn}{\nonumber}

\newcommand{\rf}{r_F}
\newcommand{\ep}{r_0^{-\gamma}}

\usepackage{hyperref}  

\usepackage{graphicx}% Include figure files
\usepackage{dcolumn}% Align table columns on decimal point
\usepackage{bm}% bold math
\begin{document}

\preprint{APS/123-QED}

\title{Holographic dictionary for Lifshitz and hyperscaling violating black holes}% 

\author{Wan Cong}

\email{wan.cong@univie.ac.at}

\affiliation{University of Vienna, Faculty of Physics, Boltzmanngasse 5, A 1090 Vienna, Austria}

\author{David Kubiz\v n\'ak}

\email{david.kubiznak@matfyz.cuni.cz}

\affiliation{Institute of Theoretical Physics, Faculty of Mathematics and Physics,
Charles University, Prague, V Hole{\v s}ovi{\v c}k{\' a}ch 2, 180 00 Prague 8, Czech Republic}

\author{Robert B. Mann}

\email{rbmann@uwaterloo.ca}

\affiliation{Department of Physics and Astronomy, University of Waterloo, Waterloo, Ontario, N2L 3G1, Canada}

\author{Manus R. Visser}

\email{manus.visser@ru.nl}

\affiliation{Institute for Mathematics, Astrophysics and Particle Physics, and Radboud Center for Natural Philosophy, \\Radboud University, 6525 AJ Nijmegen, The Netherlands}

\date{\today}% It is always \today, today,

\begin{abstract}
We develop a novel holographic dictionary for the thermodynamics of black holes with Lifshitz and hyperscaling violating asymptotics, generalizing the     dictionary for Anti-de Sitter black holes. Using our  dictionary we show that the holographic Euler equation is dual to a generalized Smarr formula for these black holes, and we  
find a precise match between the extended bulk  and   boundary first law. 
Notably, the dictionary for the central charge appearing in the Euler relation  depends on the hyperscaling violating parameter,  but not on the Lifshitz dynamical exponent.  
\end{abstract}

 \maketitle

\textbf{Introduction.} Holography, or gauge/gravity duality, is the idea   that  a theory of  quantum gravity in a ($d+1$)-dimensional bulk spacetime is dual to a quantum theory without gravity living on the $d$-dimensional asymptotic boundary of the bulk spacetime \cite{tHooft:1993dmi,Susskind:1994vu}. This duality is best understood for gravitational theories in asymptotically Anti-de Sitter (AdS) spacetime, for which the dual theory is  a  conformal field theory (CFT), also known as the `AdS/CFT correspondence' \cite{Maldacena:1997re,Gubser:1998bc,Witten:1998qj}. An important feature of this correspondence is that the thermodynamics of stationary AdS black holes \cite{Hawking:1982dh} is dual  to the thermodynamics of high-energy 
equilibrium states in a strongly coupled, large-$N$ conformal gauge theory \cite{Witten:1998zw}.
 For instance,  
 the black hole entropy $S=A/4G$ \cite{Bekenstein:1973ur,Hawking:1975vcx} is identified with the thermodynamic entropy $\tilde S$ of the CFT equilibrium state, and the internal energy $\tilde E$, temperature $\tilde T$, electric potential  $\tilde \Phi $, and  charge $\tilde Q$ of the CFT state are proportional to, respectively,  the  mass $M$,   temperature $T$, electric potential $\Phi$ and   charge $Q$ of the black hole \cite{Karch:2015rpa,Visser:2021eqk,Cong:2021jgb,Ahmed:2023snm} 
 \begin{align} \label{dictionaryAdS}
  \tilde  E = \frac{M L}{R}\,, \, \,  \tilde T = \frac{T L}{R}\,,\, \, 
  \tilde S = \frac{A}{4G}  \,,\, \,  
    \tilde \Phi =\frac{\Phi}{R}\,,\, \, 
    \tilde Q = Q L\,, 
\end{align}
where $R$ is the  curvature radius of the spatial geometry on which the CFT lives, $L$ is the AdS  curvature radius, $A$ is the black hole horizon area, and $G$ is Newton's constant.  
The proportionality factors    in the holographic dictionary for $\tilde E, \tilde T$ and $\tilde \Phi$ appear due to a specific choice of CFT metric, for which  the CFT time is $R/L$ times the   static AdS time and the   curvature radius is $R$ \cite{Gubser:1998bc,Witten:1998qj,Savonije:2001nd}.

  The general principles behind holography are expected to apply to   nonasymptotically AdS spacetimes as well. Since the isometries of the bulk spacetime should match with the global symmetries of the dual boundary theory,      field theories that are not conformally (or scale) invariant are not  dual to AdS spacetime, but possibly to a different bulk geometry.  For example, some   quantum field theories  exhibit the anisotropic scaling symmetry $\{ t, x^i\} \to \{ \alpha^z t, \alpha  x^i\}$, parametrized by the dynamical critical exponent $z$. Together with time and space translations and space rotations, this scaling symmetry forms the Lifshitz symmetries. For $z \neq 1$, Lifshitz invariant theories describe nonrelativistic quantum critical systems.         
The anisotropic scaling symmetry can be  geometrically realized in the bulk by  the  so-called Lifshitz spacetimes, which are a generalization of AdS spacetime  \cite{Son:2008ye,Balasubramanian:2008dm,Kachru:2008yh,Taylor:2008tg} (see \cite{Taylor:2015glc} for a review). Another generalization  is given by hyperscaling violating geometries, which are dual to theories violating the hyperscaling property that entropy (or free energy) scales with its naive dimension. For hyperscaling violating Lifshitz (HVL) theories (on the plane), entropy scales  with temperature as   $S \propto T^{(d-1- \theta)/z}$, where $\theta $ is the hyperscaling violation exponent \cite{Gouteraux:2011ce,Huijse:2011ef}.
Various   black hole solutions with Lifshitz and/or hyperscaling violating asymptotics have been obtained in \cite{Taylor:2008tg,Bertoldi:2009vn,Mann:2009yx,Balasubramanian:2009rx,Ayon-Beato:2009rgu,Charmousis:2010zz,Tarrio:2011de,Dong:2012se,Alishahiha:2012qu,Gouteraux:2012yr,Gath:2012pg,Bueno:2012vx,Pedraza:2018eey}.

In this note we develop   a novel   holographic   dictionary for  black hole thermodynamics in HVL theories. We find a generalization of the standard central charge dictionary $C \propto L^{d-1}/G$ for Einstein gravity, and obtain a dual relation between a generalized Smarr formula, relating the thermodynamic quantities in the bulk, and the holographic Euler relation on the boundary  \cite{Visser:2021eqk}
\begin{equation} 
\label{euler}
  \tilde   E = \tilde T \tilde S + \tilde  \Phi \tilde Q + \mu C\,.
\end{equation}
The central charge $C$ is defined as the dimensionless proportionality factor of the energy, entropy and electric charge in the canonical ensemble, i.e., 
% $)\tilde E, \tilde S, \tilde Q \propto C$
$\tilde \chi = C f_{\tilde\chi}( \tilde T, \tilde Q, V)$ for $\tilde\chi\in\{\tilde E,\tilde S,\tilde Q\}$. Furthermore, the chemical potential is defined as the conjugate quantity to $C$ in the first law,  as $\mu \equiv ( \partial \tilde E/ \partial C )_{\tilde S, \tilde Q,V}.$ We suppress the tildes on $\mu,C$ since they do not have an equivalent definition in the bulk. 
The Euler relation \eqref{euler} follows from the nonextensive scaling relation $\tilde E( \alpha  \tilde S, V,\alpha \tilde Q, \alpha C)= \alpha \tilde E(   \tilde S, V, \tilde Q,  C)$, where $V$ is the spatial volume. This scaling relation only holds in the large-$C$ limit, since there are $1/C$ corrections at finite central charge. More specifically, the Euler relation   is valid for any large-$C$ field theory at finite temperature in the deconfined phase. Moreover, it is the same for theories with different global symmetries, e.g., conformal and Lifshitz symmetries.

Furthermore, we show that an extended version of the   first law for these black holes is dual to the thermodynamic first law on the boundary
\begin{equation}
    d \tilde E = \tilde T d \tilde S + \tilde \Phi d \tilde Q - p dV +\mu dC \,,
\end{equation}
provided that $\mu$ obeys the holographic Euler equation \eqref{euler} and the boundary pressure $p$ satisfies the equation of state for Lifshitz and hyperscaling violating theories, see Eq.~\eqref{eqnofstate} below. 
 Although we work
within the context of a large class of  static, charged analytic black hole solutions to a generalized Einstein-Maxwell-Dilaton (EMD) theory~\cite{Pedraza:2018eey}, we expect that our results apply to HVL black holes in general, thus showing that  the  recently developed holographic description of extended black hole thermodynamics  in \cite{Visser:2021eqk,Ahmed:2023snm} has a wide   range of applicability.  \\

\noindent \textbf{Lifshitz and hyperscaling violating black holes.}  
A line element of the general form\footnote{Note  that we are working with $d+1$ bulk spacetime dimensions, whereas \cite{Pedraza:2018eey} has $d+2$ bulk dimensions. Moreover, we redefined $V_0$ in \cite{Pedraza:2018eey} as $-2 \Lambda_0$,  and  $\phi_0$   as $-\gamma \log r_0$, see \eqref{phi}.}
\begin{align}  
	ds^2 &= \chi \bigg [ - \left(\frac{r}{L} \right)^{2z} f(r) dt^2 + \frac{L^2}{f(r)r^2} dr^2 
 + r^2 d \Omega_{k,d-1}^2 \bigg]\,,  
 \nonumber
 \\
 \chi(r)&:= \left( \frac{r}{r_F} \right)^{-\frac{ 2 \theta}{d-1}}\,,
 \label{eq:blackholemetric}  
\end{align}
in $(d+1)$ dimensions, with $f(r) \rightarrow 1$ as $r\rightarrow \infty$, has been suggested to be dual to states in  Lifshitz invariant field theories with hyperscaling violation. For $z=1$ and $\theta=0$ we recover   asymptotically AdS spacetime. Here, $k$   parametrizes the   topology of the $(t,r)=$ constant surfaces:   $k=\{-1,0,1\}$ for   hyperbolic, planar and   spherical topology respectively.  When $f(r)=1$  under the rescalings
\begin{align}
	&t \to \alpha^z t \,, \qquad \quad \Omega \to \alpha \Omega\,, \qquad \quad 
 r \to \alpha^{-1} r \,,
 \label{scaling1}
\end{align}
the line element~\eqref{eq:blackholemetric} scales as
\begin{align}
	 ds \to \alpha^{  \theta / (d-1)} ds \,,
 \label{scaling1b}
\end{align} 
with $\alpha$ being a dimensionless scaling parameter.   
Leaving aside the   scaling of the bulk radius $r$, this rescaling   of the boundary metric is characteristic of hyperscaling violating  field theories.
Hence, we  refer to \eqref{eq:blackholemetric} as asymptotically Lifshitz and hyperscaling violating geometries (although, strictly speaking, this is only true for   $k=0$, since geometries with $k=\pm 1$ are not invariant under  spatial translations).    As is standard for hyperscaling violating theories, the dual black hole geometry is considered to be valid up to a certain  large radius $r_F$, which corresponds to a UV cutoff scale on the boundary, since the dual field theory could have a different UV completion \cite{Dong:2012se,Gath:2012pg,Perlmutter:2012he}.

 Static,  charged black holes with line element~\eqref{eq:blackholemetric}  were found as analytic solutions to a generalized Einstein-Maxwell-Dilaton  theory \cite{Pedraza:2018eey}, consisting of Einstein gravity minimally coupled to a real scalar field $\phi$, called the ``dilaton,'' and three
Abelian gauge fields $\textsf{A},\textsf{B}$ and $\textsf{C}$.  The action of the generalized EMD theory  is given by
 \begin{align}
     I  &= \frac{1}{16 \pi G} \int d^{d+1} x \sqrt{-g} \Big [ R - \frac{1}{2} (\nabla \phi)^2 + \mathcal V(\phi)  \nonumber
    \\
    &-  \frac{1}{4} X(\phi) F^2 - \frac{1}{4} Y(\phi) H^2 - \frac{1}{4} Z(\phi) K^2 \Big] \label{eq:actionemd}\,,
     \end{align}
with field strengths $F=d\textsf{A}, H = d\textsf{B}$ and $K = d\textsf{C}$ and
\begin{equation*} \label{potential}
	\mathcal{V} = -2\Lambda_0 e^{\lambda_0 \phi}\,, \ 
X = X_0 e^{\lambda_1 \phi}\,, \ 
Y= Y_0 e^{\lambda_2\phi}\,, 
 \ Z = Z_0 e^{\lambda_3 \phi}\,,  
\end{equation*}
where the $\lambda_i$ are given in terms of ($d,\theta,z)$; namely,
 \begin{align}
 	 	\lambda_0 &= \frac{2 \theta}{\gamma (d-1)}\,,
\quad
        \lambda_1=-\frac{2(d-\theta-1+\theta/(d-1))}{\gamma}\,,\nonumber\\
        \lambda_2=&-\frac{2(d-2)(d-\theta-1)}{\gamma(d-1)}\,,\quad \lambda_3=\frac{\gamma}{d-\theta -1}\,,\nonumber\\
        \gamma  =& \sqrt{2(d - \theta -1)\left(z - 1 - \frac{\theta}{ d-1 }\right )}\,,
    \end{align}
 allowing for a nonzero hyperscaling violating parameter $\theta$.    
   The gauge field $\textsf{A}$ is responsible for the anisotropic   scaling symmetry, $\textsf{B}$ supports nontrivial (spherical or hyperbolic) horizon  topology, and $\textsf{C}$ allows for nonzero electric charge.  
The analytic solution for these black holes is given by \eqref{eq:blackholemetric} with  \cite{Pedraza:2018eey}  
 \begin{align}
 	f (r) &= 1 + k \frac{(d-2)^2}{(d- \theta + z - 3)^2} \frac{L^2}{r^2} - \frac{m}{r^{d- \theta + z-1}} 
  \nonumber
  \\
  &\quad
  + \frac{q^2}{r^{2(d - \theta + z - 2)}}\,, 
  \\
  \textsf{C}&=   -\frac{\rho_3}{d- \theta + z - 3} r_0^{ \gamma \lambda_3 }  r^{- (d - \theta + z - 3)}  dt\,, 
  \\
    \phi  &= \gamma \log (r/r_0) \;, \quad  \label{phi}
    \end{align}
generalizing  previously discovered solutions for charged  spherical black holes with Lifshitz asymptotics \cite{Tarrio:2011de} (to nonzero $\theta$) and charged black branes with arbitrary Lifshitz and hyperscaling violation parameters \cite{Alishahiha:2012qu} (to nontrivial topology).   The respective mass and charge parameters  are 
 $m$ and $q$, 
 and
    \begin{align}
     L^2 &= -\frac{(d-\theta + z- 2)(d- \theta + z-1)}{ 2\Lambda_0 r_F^{2 \theta / (d-1)}} 
     r_0^{2 \theta / (d-1)}
    % r_0^{ \gamma\lambda_0}
    \,, 
     \label{eq:L}
    \\
    \rho_3^2 &= \frac{2 q^2 (d- \theta -1)(d- \theta + z-3)}{Z_0 L^{2z} r_0^{ \gamma\lambda_3}}  r_F^{2 \theta / (d-1)} \,,
    %    \\
    %\lambda_3 &= \frac{\gamma}{d- \theta %-1}\,,
     \end{align}
with $L$    the bulk curvature radius  (analogous to the AdS curvature radius), 
and $r_0$  an arbitrary length scale.  This solution is only valid for $d- \theta + z -3>0$ and $\theta < d-1$. \\ \\  
 
\noindent \textbf{Extended bulk thermodynamics.} The thermodynamic quantities of the hyperscaling violating Lifshitz black hole solution were computed in \cite{Pedraza:2018eey} (see also, e.g., \cite{Tarrio:2011de,Alishahiha:2012qu,Brenna:2015pqa,Kiritsis:2016rcb,Kastor:2018cqc}).  
The  Hawking temperature $T$, Bekenstein-Hawking entropy $S$, ADM mass $M$, electric charge $Q$, and electric potential  $\Phi$ are\footnote{The ADM mass was obtained in \cite{Pedraza:2018eey} through the background subtraction method, so that the solution with $m=0$ has zero total gravitational energy. For hyperbolic black holes, however, this solution does not  correspond to the ground state of the grand canonical ensemble, which is given by the extremal solution with zero charge and in our conventions  hence has negative energy (whereas in \cite{Pedraza:2018eey} the ADM mass for $k=-1$ was redefined such that the extremal solution has zero energy).}
\begin{widetext}
    \begin{align}\label{tempent}
        T  &=\frac{f'(r_h)}{4\pi}\Bigl(\frac{r_h}{L}\Bigr)^{z+1}\!\!= \frac{r_h^z}{4 \pi L^{z+1}} \left [  d-\theta + z-1  + \frac{k(d-2)^2}{d - \theta + z -3} \frac{L^2}{r_h^2} - \frac{(d- \theta + z - 3)q^2}{r_h^{2 (d - \theta + z-2)}}    \right]\,,\quad
        S = \frac{\omega_{k,d-1}}{4 G} r_h^{d- \theta-1} r_F^\theta \, ,
       \nonumber\\
        M &= \frac{\omega_{k,d-1}}{16 \pi G} (d- \theta-1) m L^{-z-1} r_F^\theta\,,
        \quad
        Q = \frac{\omega_{k,d-1}}{16 \pi G} Z_0 \rho_3 L^{z-1} r_F^{\theta - 2 \theta / (d-1)} 
        \,,
        \quad
        \Phi = \frac{q r_F^{\theta/ (d-1)} r_0^{\lambda_3 \gamma /2}}{\sqrt{\frac{Z_0 (d- \theta + z- 3)}{2 (d- \theta -1)}} L^z    r_h^{d - \theta + z - 3}}\,,
    \end{align}
\end{widetext}
where $r_h$ is horizon radius, given by the largest positive root of $f(r)=0$. 
The  mass $M $ depends on the  parameters $(m, L, r_F)$;  $m$ can be expressed, in turn, in terms of $r_h$, $L$, and $q$ by solving $f(r_h, L, m , q)=0$ for $m$. 
To derive a generalized Smarr relation and    first law, we  express $M$ as a function of the thermodynamic variables $S $ and $Q $, and the analog of the  ``bulk pressure'' $P:= -\frac{\Lambda_0}{8\pi G}$ for   EMD theory \cite{Kastor:2009wy}, where $\Lambda_0$ is the ``bare'' cosmological constant appearing in the Lagrangian. Due   to the coupling of $\Lambda_0$ to the dilaton, this does not correspond to the pressure of a bulk perfect fluid as usual for the cosmological constant,  but we will continue to use $P$ as a useful theory parameter.  The result (see Supplemental Material) is that $M = M(S,Q,P)$. The mass also    depends on the   length scales $r_F$ and $r_0$, but we treat them here as fixed parameters, since they do not have a clear thermodynamic interpretation from the bulk and boundary perspective.

From the mass function (\ref{mass2}) in the Supplemental Matertial the following  scaling relation may be inferred:
\begin{align}
 M( S, Q, P ) 
    = \alpha^{2+\theta-d}M(  \alpha^{d-\theta-1}S,\alpha^{d-\theta+z-3} Q,  \alpha^{-2}P)\,,\label{scaling4}
\end{align}
which   arises from
the rescaling   $r_h\to \alpha r_h$, $L\to \alpha L$ and $q \to \alpha^{d-\theta + z-2}q$. A generalized Smarr formula can now be derived from this scaling relation by applying Euler's theorem for homogeneous functions  \cite{Kastor:2009wy} 
\begin{align}
   & (d\!-\!\theta\!-\!2) M \! =\!  (d\!-\!\theta\!-\!1) TS \! +\!2 \Theta P 
    \!+\! (d\!-\!\theta\!+\!z\!-\!3)\Phi Q\,, 
    \label{smarr3}
\end{align}
with
\begin{align} \label{defsconj}
    \Theta &:= - \frac{\partial M}{\partial P}\Big |_{S,Q} 
\end{align}
 the conjugate quantity to $P$. 
If   $\theta = 0$ and $z=1$,    $\Theta$   reduces to the expected expression $- r^d\omega_{k,d-1}/d$. Hence $\Theta$ (see equation (\ref{theta1}) in the Supplemental Matertial) is the HVL generalization of (minus) the ``thermodynamic volume'', see e.g. \cite{Kubiznak:2016qmn}.

The generalized Smarr  relation \eqref{smarr3} is new in the literature, as the scaling relation \eqref{scaling4} was   not properly derived before. In fact, in \cite{Brenna:2015pqa,Pedraza:2018eey,Romero-Figueroa:2024sac} it was assumed that the standard Smarr formula for AdS black holes (given by \eqref{smarr3} with $z=1$ and $\theta=0$) also holds for HVL  black holes. However this is clearly not the case, since the Smarr relation \eqref{smarr3} explicitly involves $z$ and $\theta$.

Next, from the function $M = M(S,Q,P)$ a  variational    relation follows straightforwardly. Viewing $r_F$ and $r_0$    as fixed,
the extended first law for charged HVL black holes reads 
\begin{equation}
  	d M = T dS + \Phi dQ - \Theta dP  \,.
   \label{eq:flaw1}
\end{equation}
Since $P$ depends via $\Lambda_0$ on $L$  due to~\eqref{eq:L}, using \eqref{smarr3} we can write the extended first law   as
    \begin{align}
    &  dM = T dS  + \Phi dQ + \!\Big((d-\theta-2) M\! - \!(d-\theta -1) T S 
  \nn   \\
    &   \!-\!(d-\theta+z-3)\Phi Q\Big)\frac{dL}{L}  
   \label{bulkFlaw}
\end{align}
  This expression is useful for finding a precise match between  the bulk and boundary first laws. \\

\noindent  \textbf{Holographic dictionary.}
We would now like to find a gauge/gravity dictionary   matching \eqref{bulkFlaw} to the thermodynamic first law of the dual field theory,  given by
\begin{equation} \label{eqboundaryfirstlaw}
    d \tilde E = \tilde T d \tilde S + \tilde \Phi d \tilde Q - p dV +\mu dC  
\end{equation}
where $\tilde E$ is the internal energy of the field theory, $\tilde S=S$ and $\tilde T$ its entropy and temperature, $\tilde\Phi$ and $\tilde Q$ its electric potential and charge, $p$ and $V$ its pressure and volume, and $\mu$ and $C$ its chemical potential and central charge~\cite{Visser:2021eqk}. We   assume that the gauge theory lives on a spatial geometry with   line element   $  R^2 d \Omega_{k,d-1}^2$, 
where $R$ is the boundary curvature scale,  so that the   spatial volume is proportional to 
\begin{equation} \label{eqV}
V \propto R^{d-1}\,.
\end{equation}
The holographic dictionary follows from requiring (i) that the    extended bulk first law \eqref{bulkFlaw} exactly matches with the boundary first law \eqref{eqboundaryfirstlaw}; (ii)   that     the chemical potential $\mu$ obeys the holographic Euler equation \eqref{euler}; and (iii) that    the pressure $p$ satisfies the equation of state for HVL theories \cite{Taylor:2015glc,Kiritsis:2016rcb}
\begin{equation} \label{eqnofstate}
    p= \frac{\tilde E}{V} \frac{z-\theta / (d-1)  }{d-1-\theta}   \,.
\end{equation}
These three conditions yield the dictionary given below in   \eqref{dictC}-\eqref{eqQ}.\footnote{We also checked that the matching between the bulk and boundary extended first laws continues to hold if $r_F$ and $r_0$ are allowed to vary in the bulk (see Supplemental Material). } 
 We note that this equation of state   is a consequence of the     scaling relation  
\begin{equation} \label{ealpharelation}
    \tilde E (\tilde S, \alpha^{d-1  - \theta } V, \tilde Q, C) = \alpha^{-z  + \frac{\theta}{d-1}} \tilde E(\tilde S, V, \tilde Q,C)\,,
\end{equation}
which follows from the anisotropic scaling symmetry and the hyperscaling violating property. Differentiating \eqref{ealpharelation} with respect to $\alpha$ and setting $\alpha =1$ yields \eqref{eqnofstate}, where the pressure is defined as $p :=-  (\partial \tilde E / \partial V )_{\tilde S, \tilde Q, C}$. We now discuss the dictionary for the various thermodynamic parameters in turn.

  First, we find   the following dictionary for the central charge 
\begin{equation} \label{dictC}
      C \propto \frac{L^{d-\theta -1} r_F^\theta}{G}    \,.
\end{equation}
For $\theta =0$ this dictionary    reduces to the well-known result for Einstein gravity. The central charge is dimensionless and should only depend on theory parameters, not on the specific details of the state, such as its temperature. Hence, the dictionary for $C$ does not involve the horizon radius $r_h$, which is related to the temperature \eqref{tempent}.
% can be read off from the proportionality constant of the grand canonical free energy $F = M - T S - \Phi Q$. Since the mass, entropy and charge all scale with the central charge, $M, S, Q \propto C$, we can   directly determine its dictionary entry by looking at, for instance,   the Bekenstein-Hawking entropy \eqref{tempent}. The central charge is dimensionless and should only depend on theory parameters, not on the specific details of the state, such as its temperature. Therefore, the dictionary for $C$ cannot involve the horizon radius $r_h$, which is related to the temperature \eqref{tempent}. Multiplying and dividing   the entropy by $L^{d-\theta-1}$, we can write
% \begin{equation}
%     S \propto C x^{d-\theta -1}\,,
% \end{equation}
% with  $x = r_h/L$, where we identify    the central charge as
% \begin{equation} \label{dictC}
%       C \propto \frac{L^{d-\theta -1} r_F^\theta}{G}    \,.
% \end{equation}
 Further, it is interesting to note that the right-hand side of  \eqref{dictC} 
depends on~$\theta$ but not $z$.
 It is well known that $\theta$ effectively lowers the number of dimensions, which is also apparent from~\eqref{dictC}. Moreover, it is quite natural that \eqref{dictC} involves $r_F$, since that corresponds to a UV cutoff in the boundary theory, on which  the number of field degrees of freedom   depends.

 We note the dictionary \eqref{dictC} is consistent with the definition of $C$ as the proportionality constant of the energy, entropy and electric  charge, i.e., $\tilde E, S, \tilde Q \propto C$. For instance,   the Bekenstein-Hawking entropy \eqref{tempent} is proportional to the central charge,
\begin{equation}
    S \propto C x^{d-\theta -1}\,,
\end{equation}
where we defined  $x \equiv r_h/L$.
Since $S \propto C$ and the entropy does
not depend on $r_0$,  
 the entry \eqref{dictC} for $C$
 likewise does not.

%The products $M L$ and $T L$ in the bulk are dimensionless, and so should match with products of boundary quantities

Second, the dictionary for the internal energy and temperature follows partly from symmetry arguments. We can make the internal energy and temperature dimensionless by simply multiplying them with the bulk curvature radius $L$. On the field theory side, this should match with products of boundary quantities that are 
(from \eqref{scaling1}) invariant under the anisotropic scaling symmetry, $\{ t,x^i\} \to \{ \alpha^z t ,  \alpha x^i \}$,  
and the   simultaneous hyperscaling transformation   $ \{ t,x^i\} \to \alpha^{- \theta / (d-1)} \{ t,x^i \}$ that undoes the rescaling \eqref{scaling1b} of the  metric at the boundary. The invariant combinations are
\begin{equation} \label{dicET}
   \tilde E R^{\frac{ z- \theta / (d-1)}{1 - \theta/(d-1)}} = M L \,, \quad \tilde T R^{\frac{ z- \theta / (d-1)}{1 - \theta/(d-1)}} = TL\,.
\end{equation}
Note that the scaling dimension  of   energy and temperature is $[\tilde E]=[\tilde T]=-[t]=z -\theta / (d-1)$, and the scaling dimension of the boundary curvature scale and spatial coordinate is $[R]=[x^i]=-1+ \theta / (d-1) $~(see Appendix~D in \cite{Kiritsis:2016rcb}), so that the lhs of \eqref{dicET} is indeed dimensionless.

Next, the dictionary entries for the electric potential and charge are more complicated, but   also  follow for a large part from dimensional analysis. We take the  boundary electric charge $\tilde Q$ to be dimensionless, as for the entropy and central charge in \eqref{eqboundaryfirstlaw};     the boundary potential $\tilde \Phi$ then has the same dimensions as     $\tilde E$ and $\tilde T$, so that $\tilde \Phi R^{\frac{ z- \theta / (d-1)}{1 - \theta/(d-1)}}$ is scale invariant. The  bulk charge scales as $Q \propto q L^{-1} r_F^{\theta - \theta/(d-1)} r_0^{1-z + \theta/(d-1)} / G$
and the bulk potential as $\Phi \propto q r_F^{ \theta/(d-1) }r_0^{z- 1 - \theta/(d-1)}/ (L^z r_h^{d-\theta + z-3})$, where the charge parameter and Newton's constant have   mass dimension $[q]=-  d + \theta - z + 2 $ and $[G]=-d+1$, respectively. We can thus make the boundary charge~$\tilde Q$ and the product $\tilde \Phi R^{\frac{ z- \theta / (d-1)}{1 - \theta/(d-1)}}$ dimensionless with the following dictionary
\begin{align} \label{eqPhi}
    \tilde\Phi R^{\frac{ z- \theta / (d-1)}{1 - \theta/(d-1)}}&=   L^{z-1} \rf^{-\frac{\theta }{d-1}} r_0^{1- z + \frac{\theta}{d-1}}   \Phi\,,\\
    \tilde Q &=  L^{2-z} \rf^{\frac{\theta }{d-1}} r_0^{  z-1 - \frac{\theta}{d-1}}Q\, ,  \label{eqQ}
\end{align}
which satisfies
\begin{equation}
     \tilde\Phi \tilde QR^{\frac{ z- \theta / (d-1)}{1 - \theta/(d-1)}} = \Phi Q L \,.
\end{equation}
Note this product is independent of $r_0$ and $r_F$. 

For  $z=1 $ and $\theta=0$ this agrees with the standard holographic dictionary \cite{Visser:2021eqk}. 
There are no $\sqrt{G}$ factors due to  
a different normalization of the action 
compared to that in \cite{Ahmed:2023snm}
(in our conventions the gauge coupling constant   depends on $G$).
We further comment that this corrects an earlier dictionary (Appendix~D in   \cite{Visser:2021eqk})  for the electric charge and potential of Lifshitz black holes, where it was proposed that $\tilde \Phi = \Phi  / R^z$ and $\tilde Q = Q L$, which is different from   the dictionary above for $\theta =0$,  i.e. $\tilde \Phi = \Phi  L^z/ R^z$ and $\tilde Q = Q L^{2-z}$. This difference   ultimately arises from the fact that in \cite{Visser:2021eqk} (following \cite{Brenna:2015pqa,Pedraza:2018eey})  it was assumed that  the AdS Smarr formula also holds for Lifshitz black holes, whereas here we find that   a more careful study of the scaling relation for the mass shows that  the correct  Smarr formula is    \eqref{smarr3}. 

We emphasize that the left-hand side of the holographic dictionary  \eqref{dicET}-\eqref{eqQ} follows from invariance under Lifshitz and hyperscaling transformations. However, in order to fix the right-hand side of these equations, we imposed that the boundary and bulk extended thermodynamics should be consistent with each other.    Specifically, we imposed the three conditions stated above \eqref{eqnofstate}.

On a {\em planar} boundary geometry (for $k=0$) there is an additional relation between the thermodynamic quantities, which is dual to an extra Smarr formula in the bulk. This allows us to replace the $\mu C$ term in the Euler relation with a $pV$ term, but this comes at the expense of additional $(z,\theta,d)$-dependent prefactors in front of the $pV$ and $\tilde \Phi \tilde Q$ terms   (that are absent for $z=1, \theta=0$), namely,
\begin{equation} 
\tilde E=\tilde T\tilde S+\frac{d-\theta+z-2}{d-\theta-1}\tilde \Phi \tilde Q-\frac{z}{z-\theta/(d-1)}pV\,.
\end{equation}
This implies that the grand canonical free energy $\tilde F = \tilde E - \tilde T \tilde S - \tilde \Phi \tilde Q$ is not equal to $-pV$ for generic $z,\theta.$ We refer to the Supplemental Material for more details.

Finally, we demonstrate that the large-$C$ Euler equation \eqref{euler} is dual to the generalized Smarr relation \eqref{smarr3}. 
First, we express  the  $\Theta P$ term in the Smarr formula in terms of the boundary energy $\tilde E$
\begin{align}
\label{thetap}
   2 \Theta P&= - 2  P \frac{ \partial M}{\partial P} \Big |_{S,Q}  = L   \frac{ \partial M}{\partial L} \Big |_{S,Q}  \nn\\
   & =R^{\frac{ z-  \theta / (d-1)}{1 - \theta/(d-1)}}  \frac{\partial \tilde E}{\partial L}\Big |_{S,Q} - \frac{\tilde E}{L} R^{\frac{   z-\theta / (d-1) }{1 - \theta/(d-1)}}\,.
\end{align}
In the first equality we used the definition of $\Theta$ \eqref{defsconj}, and in the second equality we employed the fact that $\Lambda_0$ depends on $L, r_F$ and $r_0$ via \eqref{eq:L}, but the latter two length scales are fixed. In the final equality we inserted the dictionary that relates $M$ and $\tilde E$. Subsequently,
the partial derivative can be computed by noting that the boundary energy depends on the bulk quantities as follows: 
\begin{equation}
    \tilde E = \tilde E(S, \tilde Q (L,Q), V(R),C(L)) \,,
\end{equation}
yielding
\begin{align}
\frac{\partial \tilde E}{\partial L}\Big |_{S,Q} = \frac{1}{L} \left (  (2-z) \tilde \Phi \tilde Q + (d-\theta -1) \mu C \right)
\,.\end{align}
To obtain this we used the dictionary for $\tilde Q$ \eqref{eqQ} and $C$~\eqref{dictC}. 
Inserting this back into \eqref{thetap} gives
\begin{align}
    2 \Theta P &=  R^{\frac{ z-  \theta / (d-1)}{1 - \theta/(d-1)}} \frac{1}{L} \left (  (2-z) \tilde \Phi \tilde Q + (d-\theta -1) \mu C- \tilde E \right) \,.
\end{align}
This precisely agrees with the generalized Smarr formula~\eqref{smarr3} if we insert the large-$C$ Euler equation \eqref{euler} and employ the  holographic dictionary   \eqref{dicET}, \eqref{eqPhi} and~\eqref{eqQ}. This extends previous derivations  \cite{Karch:2015rpa,Visser:2021eqk,Ahmed:2023snm} of the Smarr formula from the Euler equation to HVL black holes, and it corrects Appendix~D in \cite{Visser:2021eqk} (see  \cite{Mancilla:2024spp} for an alternative holographic interpretation of the Smarr formula).  In the Supplemental Material we give further evidence for the large-$C$ Euler equation by deriving it  from a generalized Cardy formula for $d=2$, without invoking insights from holography. \\

\noindent \textbf{Conclusion.} Using the extended black hole thermodynamics as a guiding principle, we developed a novel holographic dictionary  for thermal equilibrium states dual to HVL black holes, and checked that the first law and large-$C$ Euler equation are still valid for these states. It is remarkable that the Euler equation has such a wide range of applicability.   

For future work, it would be interesting to derive the   dictionary for the thermodynamic black hole quantities using holographic renormalization    \cite{Ross:2009ar,Mann:2011hg,Chemissany:2014xsa} and, in particular, check our formula for the central charge (which can be read off from the Casimir energy for even $d$ and $k \neq 0$). Moreover,  we would like to understand whether there exists  a   thermodynamic interpretation of the length parameters $r_F$ and $r_0$, perhaps relating these to ``Lifshitz charges'', as proposed in a simplified setting in \cite{Kiritsis:2016rcb}.
 Finally, it is likely that our new setup will lead to a discovery of so far unseen phase transitions and critical phenomena arising for   Lifshitz and hyperscaling violating theories.   \\

\textbf{Acknowledgments.}
D.K. is grateful for support from Grant No. GAČR
23-07457S   of the Czech Science Foundation and the Charles University Research Center Grant No. UNCE24/SCI/016. M.R.V. acknowledges support from the Swiss National Science Foundation Postdoc Mobility Grant No. P500PT-206877 and the Spinoza Grant of the Dutch Science Organisation (NWO). R.B.M. acknowledges support from the Natural Sciences and Engineering Research Council of Canada.

   \newpage

\begin{widetext}
\appendix
{\begin{center}
    {\bf \Large Supplemental Material}
\end{center}

\section{Mass formula, scaling relations and extended first law}
\label{suppl}

The formula for   the ADM mass as a function $M = M(S,Q,P,r_F,\ep)$ is given by 
\begin{align} 
    M &=(d-1-\theta) \frac{  \rf^\theta \omega_{k,d-1}}{16\pi G} \bigg( \frac{16\pi G P  }{(d-2+z-\theta)(d-1+z-\theta)}\bigg)^{\frac{z+1}{2}}
    \bigg(\frac{4 G S }{\omega_{k,d-1} \rf^{\theta}}\bigg)^{\frac{d-1+z-\theta}{d-1-\theta}}
    \bigg( \frac{r_0}{\rf}\bigg)^{-\frac{(z+1)\theta}{d-1}}
    \times
    \nn
    \\
    & \quad
    \times\bigg( 1 + \frac{(d-2)^2  (d-\theta +z-2) (d-\theta +z-1) k }{ 16\pi  G P (d-\theta +z-3)^2} \left(\frac{r_0}{\rf}\right)^{\frac{2 \theta }{d-1}} \left(\frac{4 G S }{\omega_{k,d-1}\rf^{\theta }}\right)^{-\frac{2}{d-\theta -1}}
    \nn
    \\
    &\qquad
    + 
    \frac{8 \pi  G Q^2 r_0^{2 (z-1)}  (d-\theta +z-2) (d-\theta +z-1)}{P Z_0 \rf^{2 \theta }(d-\theta -1) (d-\theta +z-3) \omega_{k,d-1}^2 }
    \left(\frac{4 G S }{\omega_{k,d-1}\rf^{\theta }}\right)^{-\frac{2 (d-\theta +z-2)}{d-\theta -1}} \bigg)\,.
    \label{mass2}
\end{align}
In this appendix we treat the length scales $r_0$ and $r_F$ as variable parameters, and we derive  scaling relations for the mass function that follow from rescaling $r_0$ and $r_F$.  In the main text we kept $r_0$ and $r_F$ fixed, since they have no clear thermodynamic interpretation, but it is tantalizing to think that they might have a thermodynamic meaning in an `extended' version of black hole mechanics (perhaps by varying the charges associated to the $\textsf{A}$ and $\textsf{B}$ gauge fields, see (3.13) and (3.14) in \cite{Pedraza:2018eey}). 

Using \eqref{mass2} 
it is straightforward to directly verify the following three independent scaling relations:
%{\bf to save space we could write something like:}
\begin{eqnarray}
 M( S, Q, P,  \rf,  \ep)\, &&= \alpha^{-\theta}M(\alpha^{\theta} S,\alpha^{\theta- \theta/(d-1)} Q,\alpha^{-2 \theta/ (d-1)} P,  \alpha \rf,  \ep) \quad \  \label{scaling2}
    \\
   &&= M(  S,\alpha^{\lambda_3/2} Q,  \alpha^{-\lambda_0}P,   \rf,  \alpha \ep) 
    \label{scaling3}\\    
    &&= \alpha^{2+\theta-d}M(  \alpha^{d-\theta-1}S,\alpha^{d-\theta+z-3} Q,  \alpha^{-2}P,   \rf,    \ep)\,,\label{scaling4b}
\end{eqnarray}
 which respectively arise from the rescalings  ($r_F\to \alpha r_F$, $r_0^{-\gamma} \to \alpha r_0^{-\gamma}$), and   ($r_h\to \alpha r_h$, $L\to \alpha L$ and $q \to \alpha^{d-\theta + z-2}q$). In AdS space ($z=1$, $\theta =0$) only the third scaling relation survives, from which the Smarr formula for AdS black holes is usually derived by applying Euler's theorem for homogeneous functions \cite{Kastor:2009wy}. For arbitrary $z$ and $\theta,$ however, we have two additional scaling relations due to the presence of two new   length scales $r_F$ and $r_0.$ By applying Euler's theorem to the three scaling relations \eqref{scaling2}-\eqref{scaling4b}, we thus respectively find       three independent Smarr-type formulas for HVL black holes 
\begin{align}
   & \theta M = \theta T S +  \frac{(d-2) \theta}{d-1}
   \Phi Q + U \rf +\frac{2\theta}{d-1} \Theta P\,,
 \label{smarr1}\\
&0 = \frac{\lambda_3}{2} \Phi Q + U_2 r_0^{-\gamma} + \lambda_0 \Theta P\,, 
 \label{smarr2}
\\
   & (d\!-\!\theta\!-\!2) M \! =\!  (d\!-\!\theta\!-\!1) TS \! +\!2 \Theta P 
    \!+\! (d\!-\!\theta\!+\!z\!-\!3)\Phi Q \,.
    \label{smarr3b}
\end{align}
The   conjugate variables to the ``bulk pressure'' $P:=-  \frac{\Lambda_0}{8\pi G}$ and to the length scales $r_F$ and $r_0^{-\gamma}$ are defined as partial derivatives of the mass, and are, respectively, given by 
\begin{align}
     \Theta := -\frac{\partial M}{\partial P}\Big |_{S,Q,r_F,r_0}&= -
      \frac{L^{1-z} \rf^{\frac{(d+1) \theta }{d-1}} r_0^{-\gamma  \lambda_0} r_h^{d-\theta +z-3} (d-\theta -1) \omega_{k,d-1} }{2 (d-\theta +z-3)^2 (d-\theta +z-2) (d-\theta +z-1)} 
      \times
    \label{theta1} \\
     &\quad
     \times\bigg( r_h^{-2 (d+z-1)} (d-\theta +z-3)^2 \big((z+1) r_h^{2 (d+z)}+q^2 (z-1) r_h^{2 \theta +4}\big)+(d-2)^2 k L^2 (z-1)\bigg)
    \,, \nn
    \\
    U := \frac{\partial M}{\partial r_F} \Big |_{S,Q,P,r_0}
    &= 
   -\frac{\theta  \omega_{k,d-1} L^{-z-1} \rf^{\theta -1} r_h^{-d-\theta -z-3}}{16 \pi G (d-1)  (d-\theta +z-3)^2}
   \bigg[ r_h^2 (d-\theta +z-3)^2 \bigg( q^2 r_h^{2 \theta +4} ((1-2z) d +\theta +(\theta +2) z-1)
    \nonumber
    \\
    &\quad
    +r_h^{2 (d+z)} ((1 + z) \theta +1 -d ) \bigg) 
    -(d-2)^2 k L^2 r_h^{2 (d+z)} (d+(1-z)\theta-1) \bigg]
    \,,
    \\
    U_2:= \frac{\partial M}{\partial (\ep)} \Big |_{S,Q,P,r_F}
    %\big|_{S,Q,L,\rf,G} 
    &= 
    \frac{(d-\theta -1) \omega_{k,d-1} L^{-z-1} r_0^{\gamma } \rf^{\theta } r_h^{-d-\theta -z-3}}{32 \pi  G (d-\theta +z-3)^2}
    \bigg[ (d-2)^2  \lambda_0 k L^2 (z-1) r_h^{2 (d+z)}
    \nonumber
    \\
    &\quad
    +r_h^2 (d-\theta +z-3)^2 \left(\lambda_0 (z+1) r_h^{2 (d+z)}+q^2 r_h^{2 \theta +4} (\lambda_0 (z-1)-2 \lambda_3)\right)
    \bigg]
   \,.
\end{align}
The quantity $U$ vanishes if $\theta = 0$, and  $U_2$ is zero if both $\theta = 0$ and $z=1$, in which case   $\Theta$   reduces to  $- r^d\omega_{k,d-1}/d$.

  It is possible to get rid of the explicit $z$ and $\theta$ dependence in the Smarr relation, but this comes at the expense of introducing two additional terms, proportional to $r_F$ and $r_0^{-\gamma}$, yielding
the generalized Smarr formula 
\begin{equation}
\label{eq:smarr2}
   (d-2) M = (d-1) TS + (d-2)\Phi Q+ 2 \Theta P + U \rf - \gamma U_2\ep\,,
\end{equation}
obtained by combining \eqref{smarr1}-\eqref{smarr3b}.
Note that the prefactors are the same as for the Smarr formula for AdS black holes \cite{Kastor:2009wy}. Indeed, for $(\theta =0$, $z=1)$, $U=U_2=0$,  reproducing the AdS Smarr formula. The Smarr relation  \eqref{eq:smarr2} also follows from the scaling relation
 \begin{align} &M(\alpha^{d-1} S,\alpha^{d-2} Q,\alpha^{-2 (1 + \theta/(d-1))} P,  \alpha \rf, \alpha^{-\gamma}\ep)   = \alpha^{d-2} M( S, Q, P,  \rf,  \ep) \,,
 \end{align}
which is a combination of the   scaling relations \eqref{scaling2}-\eqref{scaling4b}.

Next, we  use the function $M = M(S,Q,P, r_F, r_0^{-\gamma})$ to derive an extended version of the first law for charged HVL black holes, that includes variations of $r_F$ and $r_0^{-\gamma},$   
\begin{equation}
  	d M = T dS + \Phi dQ - \Theta dP + U d\rf + U_2 d r_0^{-\gamma} \,.
   \label{eq:flaw1b}
\end{equation}Since the pressure depends via $\Lambda_0$ on $L$, $r_F$ and $r_0$, due to~(12) in the main text, the extended first law can also be written in terms of variations of these length scales
    \begin{align}
    &dM = T dS  + \Phi dQ + \big(\theta M - \theta  T S -\frac{(d-2) \theta}{d-1}
   \Phi Q\big) \frac{d\rf}{\rf}    \nn \\
    & + \!\big((d-\theta-2) M\! - \!(d-\theta -1) T S \!-\!(d-\theta+z-3)\Phi Q\big)\frac{dL}{L} +\left ( z-1 - \frac{\theta}{d-1} \right) \Phi Q \frac{dr_0}{r_0} \,,
   \label{bulkFlawb}
\end{align}
using \eqref{smarr1}-\eqref{smarr3b} to replace $\Theta$, $U$ and $U_2$ with the   thermodynamic variables $M,T,S,\Phi$ and $Q$. We checked  this matches with the extended boundary first law (20) if the pressure satisfies the HVL equation of state and the chemical potential obeys the holographic Euler equation. The match  follows from the same holographic dictionary for the thermodynamic variables as used in the main text,   (21) and (23)-(26). Hence, this strengthens the holographic dictionary derived in the main text. We do note, however, that the boundary theory seems to organize the thermodynamic variables in a more efficient way, since the boundary energy depends on four other variables, $\tilde E = \tilde E(\tilde S, \tilde Q, V, C)$, whereas the bulk mass  seems to depend on five variables $M = M(S, Q, P, r_F, \ep)$ or $M= M (r_h, q, L, r_F, \ep)$. Thus, there seems to be a certain redundancy in the bulk thermodynamic description.

\section{Smarr formula and Euler equation for planar geometries}

For HVL black branes (with planar horizon topology,  $k=0$) the mass satisfies one additional, independent scaling relation 
\begin{align} \label{newscaling}
    &M (\alpha^{d-\theta-1}S,\alpha^{d-\theta + z-2}Q, P, r_F, r_0^{-\gamma}) = \alpha^{d-\theta + z-1}M ( S, Q, P, r_F, r_0^{-\gamma}) \,,
\end{align}
from which the following Smarr relation can be derived \cite{Brenna:2015pqa,Pedraza:2018eey}
\begin{equation} \label{addiationalsmarr}
    (d-\theta + z-1) M = (d-\theta -1)TS + (d-\theta + z-2)\Phi Q\,.
\end{equation}
Together with \eqref{smarr3b}  this implies
\begin{equation}
    (z+1) M = - 2 \Theta P + \Phi Q \,.
\end{equation}
This Smarr-like formula can, in turn, be derived from the scaling relation 
\begin{align} \label{blascaling}
    &M ( S,\alpha^{-1}Q, \alpha^{-2}P, r_F, r_0^{-\gamma}) = \alpha^{- (z+1)}M ( S, Q, P, r_F, r_0^{-\gamma}) \,.
\end{align}
This scaling relation follows from combining \eqref{scaling4b}  and \eqref{newscaling}, so there is only one new scaling relation in addition to the three scaling relations in the previous appendix.
The scaling relation  \eqref{newscaling} arises from rescaling $r_h \to \alpha r_h$ and $q \to \alpha^{d-\theta + z-2} q$, while \eqref{blascaling} follows from sending $L \to \alpha L$.

In the holographic field theory,  the thermal state   that is dual to the black brane lives on the plane. Here as well there are extra scaling relations for the thermodynamic quantities, namely
\begin{equation}
    \tilde E \propto \tilde  T^{\frac{d-\theta + z-1}{z}}\, , \quad \tilde S \propto  \tilde T^{\frac{d-\theta-1}{z}}\, ,\quad \tilde Q \propto \tilde T^{\frac{d-\theta + z-2}{z}}\,.
\end{equation}
This implies   the additional scaling relation
\begin{align} \label{newscalingb}
    &\tilde E (\alpha^{d-\theta-1} \tilde S,\alpha^{d-\theta + z-2}\tilde Q, V, C)  = \alpha^{d-\theta + z-1}\tilde E ( \tilde S,\tilde  Q,  V, C) \,,
\end{align}
which yields
\begin{equation} \label{cftextrarelation}
    (d-\theta + z-1) \tilde E = (d-\theta -1) \tilde T\tilde S + (d-\theta + z-2)\tilde \Phi \tilde Q\,.
\end{equation}
This is obviously dual to the additional Smarr relation \eqref{addiationalsmarr} in the bulk, upon invoking the holographic dictionary developed in the main text. 

Now, if we combine   formula \eqref{cftextrarelation} with the holographic Euler equation (2) (which is still valid on the plane) and the Lifshitz, hyperscaling violating equation of state (28), we find the relation  
\begin{equation}
    \mu C =   \frac{z-1}{d -\theta-1} \tilde \Phi \tilde Q
 - \frac{z}{z- \theta /(d-1)} pV\,. 
 \end{equation}
 In other words,   the boundary grand canonical free energy $\tilde F = \tilde E - \tilde T \tilde S - \tilde \Phi \tilde Q$ (that equals $\mu C$ due to the holographic Euler equation) is not proportional to $pV$, but contains an additional $\tilde \Phi \tilde Q$ term.  
If we insert this relation back into the holographic Euler equation, then we find an unusual thermodynamic relation
\begin{equation} \label{genereuler}
    \tilde E = \tilde T \tilde S + \frac{d-\theta + z-2}{d-\theta -1} \tilde \Phi \tilde Q - \frac{z}{z-\theta / (d-1)} pV\,.
\end{equation}
For $z=1, \theta=0$ this reduces to the standard thermodynamic Euler equation $ \tilde E = \tilde T \tilde S + \tilde \Phi \tilde Q - pV$, that is valid for a thermal CFT on a plane. Also for $\theta =0$ and $\tilde Q=0$, equation \eqref{genereuler} simplifies to the standard Euler equation $ \tilde E = \tilde T \tilde S   - pV. $ However, for arbitrary $z$, $\theta$ the generalized Euler equation that involves a $pV$ term seems to be \eqref{genereuler} (in contrast to what was claimed in Appendix C of \cite{Visser:2021eqk} for the case of $\theta =0$ and $\tilde Q \neq 0$).  

\section{Euler equation from a generalized Cardy formula}

  In $(1+1)$  %$d=2$ 
 dimensions the large-$C$ Euler equation can  be obtained   purely from the  field theory side without invoking holography. For two-dimensional CFTs it was shown in \cite{Visser:2021eqk} that  the   Euler equation follows from the Cardy formula for the microcanonical entropy \cite{Cardy:1986ie}.  Here we extend that proof to Lifshitz and hyperscaling violating theories using a generalized Cardy formula derived in \cite{Gonzalez:2011nz,Shaghoulian:2015dwa}. This generalized Cardy formula is based on a generalized notion of modular invariance, which allows one to project the partition function at high temperatures to its vacuum state. For Lifshitz theories this invariance has been criticised in \cite{Shaghoulian:2015dwa}, since Lifshitz symmetries do not allow one to swap thermal and spatial cycles because it does not contain boosts (for hyperscaling theories this is not an issue since they are Lorentz boost invariant).  However, below we proceed  with analyzing the implications of the    generalized modular invariance and Cardy formula as if it exists.   
 
 For arbitrary $(z,\theta)$   the generalized Cardy formula for vanishing angular momentum reads \cite{Gonzalez:2011nz,Shaghoulian:2015dwa,BravoGaete:2017dso}   
 \begin{equation} \label{eq:generalizedcardy}
 \tilde 	S=\frac{1 + z- \theta }{1 - \theta} 2 \pi  \tilde E R^{\frac{z - \theta }{1- \theta  }}  \left ( \frac{- \tilde E_{\text{vac}} (1- \theta) }{z \tilde E}\right)^{\frac{z}{1 - \theta + z}} \,.
 \end{equation}
We have reinstated the radius $R$ of the spatial circle on which the field theory lives, such that the energy is dimensionful and the product $ \tilde E R^{\frac{z - \theta }{1- \theta  }}$ is scale invariant. (For $\theta=0$ the scaling   $S \propto R^z$   is different from     (9) in   \cite{Gonzalez:2011nz}, where $S \propto R$, but   our formula seems to have the correct dimensions.)   Note the Cardy formula is expressed in terms of the (negative) vacuum energy $\tilde E_{\text{vac}}$ instead of the central charge as usual. We can reintroduce a (positive and dimensionless) central charge by defining it in terms of the vacuum Casimir energy
 \begin{equation} \label{eq:centralchargecasimir}
	C:=-\tilde E_{\text{vac}} R^{\frac{z- \theta  }{1- \theta }} \, .
\end{equation}
As an example, for two-dimensional CFTs we normalize  our central charge as $C=c/12$, where $c$ is the standard central charge appearing in the Casimir energy. For $z = 1$ and $\theta =0$ equation \eqref{eq:generalizedcardy} reduces to the   Cardy formula $\tilde S=4 \pi R \sqrt{- \tilde E\tilde  E_{\text{vac}}}$  or $\tilde S =4 \pi   \sqrt{\frac{c}{6}L_0}$, where we used $  \tilde E R=2L_0$ and $\tilde  E_{\text{vac}} R= - c/12$ to obtain the  latter,  more familiar Cardy  expression (for $L_0 = 
\bar L_0$).

Since the Cardy formula expresses the entropy as a function $\tilde S=\tilde S(\tilde E,V,C),$ with central charge $C$ defined in \eqref{eq:centralchargecasimir} and the ``volume'' of the circle given by $V= 2\pi R$, we can explicitly compute  the temperature, pressure and chemical potential associated with  the central charge using the definitions that follow from the first law  
\begin{align}
	\tilde T &:=\! \left (\frac{\partial \tilde  S}{\partial \tilde E} \right)_{V,C}^{-1} \! =\! \frac{1}{2 \pi R^{\frac{z-\theta}{1- \theta}}}\! \left ( \frac{-\tilde  E_{\text{vac}} (1-\theta)}{z \tilde E}\right)^{-\frac{z}{1+z-\theta}} , \label{eq:2dtemperature}\\
	p &:= T\left (\frac{\partial \tilde  S}{\partial V} \right)_{\tilde E,C}= \frac{E}{V} \frac{z - \theta}{1-\theta} \,,\\
	\mu &:= -T\left (\frac{\partial \tilde S}{\partial C} \right)_{\tilde E,V}= -\frac{\tilde E   }{C  } \frac{z}{1- \theta}\,. 
\end{align}
In terms of the temperature and central charge the generalized  Cardy formula \eqref{eq:generalizedcardy} can be written  as
 \begin{equation} \label{eq:canonical2dentropy}
 \tilde 	S=  \frac{1+z-\theta}{z} (2 \pi)^{\frac{1+z -\theta}{z}}C\left(R^{\frac{z-\theta}{1-\theta}}\tilde T \right)^{\frac{1-\theta}{z}}\,.  
 \end{equation}
Multiplying these thermodynamic variables with their respective conjugate quantities yields 
\begin{equation}
	\tilde T\tilde S =\tilde  E \frac{1+ z - \theta}{1- \theta} \,,\qquad\,\, pV= \tilde E \frac{  z - \theta}{1- \theta} \,,\qquad\, \, \mu C=- \tilde E \frac{  z  }{1- \theta}\,.
\end{equation}
The second relation is the equation of state for a two-dimensional  hyperscaling violating theory with Lifshitz symmetry. Moreover, the first and third relation together form the large-$C$ Euler equation (in the absence of charge) 
\begin{equation} \label{eq:2deuler}
	\tilde E =\tilde  T\tilde  S + \mu C \,.
\end{equation}
Hence the Cardy formula knows about the large-$C$ Euler equation! Note the Euler equation does not depend explicitly on $z$ and $\theta$. However, if we replace the $\mu C$ term by a $pV$ term then the formula does depend  explicitly on $z$ and $\theta$
(consistent with \eqref{genereuler})\begin{equation} \label{eq:2dnoteuler}
	\tilde E =\tilde  T\tilde  S - \frac{z}{z-\theta} pV\,.
\end{equation}
Therefore, the large-$C$ Euler equation \eqref{eq:2deuler} seems to be a  universal thermodynamic relation in large-$C$   field theories, in contrast to \eqref{eq:2dnoteuler}, which depends on the symmetries of the theory and the number of dimensions. 
}
\end{widetext}

 \bibliography{HSVLifschitz}

\begin{thebibliography}{48}%
\makeatletter
\providecommand \@ifxundefined [1]{%
 \@ifx{#1\undefined}
}%
\providecommand \@ifnum [1]{%
 \ifnum #1\expandafter \@firstoftwo
 \else \expandafter \@secondoftwo
 \fi
}%
\providecommand \@ifx [1]{%
 \ifx #1\expandafter \@firstoftwo
 \else \expandafter \@secondoftwo
 \fi
}%
\providecommand \natexlab [1]{#1}%
\providecommand \enquote  [1]{``#1''}%
\providecommand \bibnamefont  [1]{#1}%
\providecommand \bibfnamefont [1]{#1}%
\providecommand \citenamefont [1]{#1}%
\providecommand \href@noop [0]{\@secondoftwo}%
\providecommand \href [0]{\begingroup \@sanitize@url \@href}%
\providecommand \@href[1]{\@@startlink{#1}\@@href}%
\providecommand \@@href[1]{\endgroup#1\@@endlink}%
\providecommand \@sanitize@url [0]{\catcode `\\12\catcode `\$12\catcode
  `\&12\catcode `\#12\catcode `\^12\catcode `\_12\catcode `\%12\relax}%
\providecommand \@@startlink[1]{}%
\providecommand \@@endlink[0]{}%
\providecommand \url  [0]{\begingroup\@sanitize@url \@url }%
\providecommand \@url [1]{\endgroup\@href {#1}{\urlprefix }}%
\providecommand \urlprefix  [0]{URL }%
\providecommand \Eprint [0]{\href }%
\providecommand \doibase [0]{https://doi.org/}%
\providecommand \selectlanguage [0]{\@gobble}%
\providecommand \bibinfo  [0]{\@secondoftwo}%
\providecommand \bibfield  [0]{\@secondoftwo}%
\providecommand \translation [1]{[#1]}%
\providecommand \BibitemOpen [0]{}%
\providecommand \bibitemStop [0]{}%
\providecommand \bibitemNoStop [0]{.\EOS\space}%
\providecommand \EOS [0]{\spacefactor3000\relax}%
\providecommand \BibitemShut  [1]{\csname bibitem#1\endcsname}%
\let\auto@bib@innerbib\@empty
%</preamble>
\bibitem [{\citenamefont {'t~Hooft}(1993)}]{tHooft:1993dmi}%
  \BibitemOpen
  \bibfield  {author} {\bibinfo {author} {\bibfnamefont {G.}~\bibnamefont
  {'t~Hooft}},\ }\bibfield  {title} {\bibinfo {title} {{Dimensional reduction
  in quantum gravity}},\ }\href@noop {} {\bibfield  {journal} {\bibinfo
  {journal} {Conf. Proc. C}\ }\textbf {\bibinfo {volume} {930308}},\ \bibinfo
  {pages} {284} (\bibinfo {year} {1993})},\ \Eprint
  {https://arxiv.org/abs/gr-qc/9310026} {arXiv:gr-qc/9310026} \BibitemShut
  {NoStop}%
\bibitem [{\citenamefont {Susskind}(1995)}]{Susskind:1994vu}%
  \BibitemOpen
  \bibfield  {author} {\bibinfo {author} {\bibfnamefont {L.}~\bibnamefont
  {Susskind}},\ }\bibfield  {title} {\bibinfo {title} {{The World as a
  hologram}},\ }\href {https://doi.org/10.1063/1.531249} {\bibfield  {journal}
  {\bibinfo  {journal} {J. Math. Phys.}\ }\textbf {\bibinfo {volume} {36}},\
  \bibinfo {pages} {6377} (\bibinfo {year} {1995})},\ \Eprint
  {https://arxiv.org/abs/hep-th/9409089} {arXiv:hep-th/9409089} \BibitemShut
  {NoStop}%
\bibitem [{\citenamefont {Maldacena}(1998)}]{Maldacena:1997re}%
  \BibitemOpen
  \bibfield  {author} {\bibinfo {author} {\bibfnamefont {J.~M.}\ \bibnamefont
  {Maldacena}},\ }\bibfield  {title} {\bibinfo {title} {{The Large N limit of
  superconformal field theories and supergravity}},\ }\href
  {https://doi.org/10.4310/ATMP.1998.v2.n2.a1} {\bibfield  {journal} {\bibinfo
  {journal} {Adv. Theor. Math. Phys.}\ }\textbf {\bibinfo {volume} {2}},\
  \bibinfo {pages} {231} (\bibinfo {year} {1998})},\ \Eprint
  {https://arxiv.org/abs/hep-th/9711200} {arXiv:hep-th/9711200} \BibitemShut
  {NoStop}%
\bibitem [{\citenamefont {Gubser}\ \emph {et~al.}(1998)\citenamefont {Gubser},
  \citenamefont {Klebanov},\ and\ \citenamefont {Polyakov}}]{Gubser:1998bc}%
  \BibitemOpen
  \bibfield  {author} {\bibinfo {author} {\bibfnamefont {S.~S.}\ \bibnamefont
  {Gubser}}, \bibinfo {author} {\bibfnamefont {I.~R.}\ \bibnamefont
  {Klebanov}},\ and\ \bibinfo {author} {\bibfnamefont {A.~M.}\ \bibnamefont
  {Polyakov}},\ }\bibfield  {title} {\bibinfo {title} {{Gauge theory
  correlators from noncritical string theory}},\ }\href
  {https://doi.org/10.1016/S0370-2693(98)00377-3} {\bibfield  {journal}
  {\bibinfo  {journal} {Phys. Lett. B}\ }\textbf {\bibinfo {volume} {428}},\
  \bibinfo {pages} {105} (\bibinfo {year} {1998})},\ \Eprint
  {https://arxiv.org/abs/hep-th/9802109} {arXiv:hep-th/9802109} \BibitemShut
  {NoStop}%
\bibitem [{\citenamefont {Witten}(1998{\natexlab{a}})}]{Witten:1998qj}%
  \BibitemOpen
  \bibfield  {author} {\bibinfo {author} {\bibfnamefont {E.}~\bibnamefont
  {Witten}},\ }\bibfield  {title} {\bibinfo {title} {{Anti-de Sitter space and
  holography}},\ }\href {https://doi.org/10.4310/ATMP.1998.v2.n2.a2} {\bibfield
   {journal} {\bibinfo  {journal} {Adv. Theor. Math. Phys.}\ }\textbf {\bibinfo
  {volume} {2}},\ \bibinfo {pages} {253} (\bibinfo {year}
  {1998}{\natexlab{a}})},\ \Eprint {https://arxiv.org/abs/hep-th/9802150}
  {arXiv:hep-th/9802150} \BibitemShut {NoStop}%
\bibitem [{\citenamefont {Hawking}\ and\ \citenamefont
  {Page}(1983)}]{Hawking:1982dh}%
  \BibitemOpen
  \bibfield  {author} {\bibinfo {author} {\bibfnamefont {S.~W.}\ \bibnamefont
  {Hawking}}\ and\ \bibinfo {author} {\bibfnamefont {D.~N.}\ \bibnamefont
  {Page}},\ }\bibfield  {title} {\bibinfo {title} {{Thermodynamics of Black
  Holes in anti-De Sitter Space}},\ }\href {https://doi.org/10.1007/BF01208266}
  {\bibfield  {journal} {\bibinfo  {journal} {Commun. Math. Phys.}\ }\textbf
  {\bibinfo {volume} {87}},\ \bibinfo {pages} {577} (\bibinfo {year}
  {1983})}\BibitemShut {NoStop}%
\bibitem [{\citenamefont {Witten}(1998{\natexlab{b}})}]{Witten:1998zw}%
  \BibitemOpen
  \bibfield  {author} {\bibinfo {author} {\bibfnamefont {E.}~\bibnamefont
  {Witten}},\ }\bibfield  {title} {\bibinfo {title} {{Anti-de Sitter space,
  thermal phase transition, and confinement in gauge theories}},\ }\href
  {https://doi.org/10.4310/ATMP.1998.v2.n3.a3} {\bibfield  {journal} {\bibinfo
  {journal} {Adv. Theor. Math. Phys.}\ }\textbf {\bibinfo {volume} {2}},\
  \bibinfo {pages} {505} (\bibinfo {year} {1998}{\natexlab{b}})},\ \Eprint
  {https://arxiv.org/abs/hep-th/9803131} {arXiv:hep-th/9803131} \BibitemShut
  {NoStop}%
\bibitem [{\citenamefont {Bekenstein}(1973)}]{Bekenstein:1973ur}%
  \BibitemOpen
  \bibfield  {author} {\bibinfo {author} {\bibfnamefont {J.~D.}\ \bibnamefont
  {Bekenstein}},\ }\bibfield  {title} {\bibinfo {title} {{Black holes and
  entropy}},\ }\href {https://doi.org/10.1103/PhysRevD.7.2333} {\bibfield
  {journal} {\bibinfo  {journal} {Phys. Rev. D}\ }\textbf {\bibinfo {volume}
  {7}},\ \bibinfo {pages} {2333} (\bibinfo {year} {1973})}\BibitemShut
  {NoStop}%
\bibitem [{\citenamefont {Hawking}(1975)}]{Hawking:1975vcx}%
  \BibitemOpen
  \bibfield  {author} {\bibinfo {author} {\bibfnamefont {S.~W.}\ \bibnamefont
  {Hawking}},\ }\bibfield  {title} {\bibinfo {title} {{Particle Creation by
  Black Holes}},\ }\href {https://doi.org/10.1007/BF02345020} {\bibfield
  {journal} {\bibinfo  {journal} {Commun. Math. Phys.}\ }\textbf {\bibinfo
  {volume} {43}},\ \bibinfo {pages} {199} (\bibinfo {year} {1975})},\ \bibinfo
  {note} {[Erratum: Commun.Math.Phys. 46, 206 (1976)]}\BibitemShut {NoStop}%
\bibitem [{\citenamefont {Karch}\ and\ \citenamefont
  {Robinson}(2015)}]{Karch:2015rpa}%
  \BibitemOpen
  \bibfield  {author} {\bibinfo {author} {\bibfnamefont {A.}~\bibnamefont
  {Karch}}\ and\ \bibinfo {author} {\bibfnamefont {B.}~\bibnamefont
  {Robinson}},\ }\bibfield  {title} {\bibinfo {title} {{Holographic Black Hole
  Chemistry}},\ }\href {https://doi.org/10.1007/JHEP12(2015)073} {\bibfield
  {journal} {\bibinfo  {journal} {JHEP}\ }\textbf {\bibinfo {volume} {12}},\
  \bibinfo {pages} {073}},\ \Eprint {https://arxiv.org/abs/1510.02472}
  {arXiv:1510.02472 [hep-th]} \BibitemShut {NoStop}%
\bibitem [{\citenamefont {Visser}(2022)}]{Visser:2021eqk}%
  \BibitemOpen
  \bibfield  {author} {\bibinfo {author} {\bibfnamefont {M.~R.}\ \bibnamefont
  {Visser}},\ }\bibfield  {title} {\bibinfo {title} {{Holographic
  thermodynamics requires a chemical potential for color}},\ }\href
  {https://doi.org/10.1103/PhysRevD.105.106014} {\bibfield  {journal} {\bibinfo
   {journal} {Phys. Rev. D}\ }\textbf {\bibinfo {volume} {105}},\ \bibinfo
  {pages} {106014} (\bibinfo {year} {2022})},\ \Eprint
  {https://arxiv.org/abs/2101.04145} {arXiv:2101.04145 [hep-th]} \BibitemShut
  {NoStop}%
\bibitem [{\citenamefont {Cong}\ \emph {et~al.}(2022)\citenamefont {Cong},
  \citenamefont {Kubiznak}, \citenamefont {Mann},\ and\ \citenamefont
  {Visser}}]{Cong:2021jgb}%
  \BibitemOpen
  \bibfield  {author} {\bibinfo {author} {\bibfnamefont {W.}~\bibnamefont
  {Cong}}, \bibinfo {author} {\bibfnamefont {D.}~\bibnamefont {Kubiznak}},
  \bibinfo {author} {\bibfnamefont {R.~B.}\ \bibnamefont {Mann}},\ and\
  \bibinfo {author} {\bibfnamefont {M.~R.}\ \bibnamefont {Visser}},\ }\bibfield
   {title} {\bibinfo {title} {{Holographic CFT phase transitions and
  criticality for charged AdS black holes}},\ }\href
  {https://doi.org/10.1007/JHEP08(2022)174} {\bibfield  {journal} {\bibinfo
  {journal} {JHEP}\ }\textbf {\bibinfo {volume} {08}},\ \bibinfo {pages}
  {174}},\ \Eprint {https://arxiv.org/abs/2112.14848} {arXiv:2112.14848
  [hep-th]} \BibitemShut {NoStop}%
\bibitem [{\citenamefont {Ahmed}\ \emph {et~al.}(2023)\citenamefont {Ahmed},
  \citenamefont {Cong}, \citenamefont {Kubiz\v{n}\'ak}, \citenamefont {Mann},\
  and\ \citenamefont {Visser}}]{Ahmed:2023snm}%
  \BibitemOpen
  \bibfield  {author} {\bibinfo {author} {\bibfnamefont {M.~B.}\ \bibnamefont
  {Ahmed}}, \bibinfo {author} {\bibfnamefont {W.}~\bibnamefont {Cong}},
  \bibinfo {author} {\bibfnamefont {D.}~\bibnamefont {Kubiz\v{n}\'ak}},
  \bibinfo {author} {\bibfnamefont {R.~B.}\ \bibnamefont {Mann}},\ and\
  \bibinfo {author} {\bibfnamefont {M.~R.}\ \bibnamefont {Visser}},\ }\bibfield
   {title} {\bibinfo {title} {{Holographic Dual of Extended Black Hole
  Thermodynamics}},\ }\href {https://doi.org/10.1103/PhysRevLett.130.181401}
  {\bibfield  {journal} {\bibinfo  {journal} {Phys. Rev. Lett.}\ }\textbf
  {\bibinfo {volume} {130}},\ \bibinfo {pages} {181401} (\bibinfo {year}
  {2023})},\ \Eprint {https://arxiv.org/abs/2302.08163} {arXiv:2302.08163
  [hep-th]} \BibitemShut {NoStop}%
\bibitem [{\citenamefont {Savonije}\ and\ \citenamefont
  {Verlinde}(2001)}]{Savonije:2001nd}%
  \BibitemOpen
  \bibfield  {author} {\bibinfo {author} {\bibfnamefont {I.}~\bibnamefont
  {Savonije}}\ and\ \bibinfo {author} {\bibfnamefont {E.~P.}\ \bibnamefont
  {Verlinde}},\ }\bibfield  {title} {\bibinfo {title} {{CFT and entropy on the
  brane}},\ }\href {https://doi.org/10.1016/S0370-2693(01)00467-1} {\bibfield
  {journal} {\bibinfo  {journal} {Phys. Lett. B}\ }\textbf {\bibinfo {volume}
  {507}},\ \bibinfo {pages} {305} (\bibinfo {year} {2001})},\ \Eprint
  {https://arxiv.org/abs/hep-th/0102042} {arXiv:hep-th/0102042} \BibitemShut
  {NoStop}%
\bibitem [{\citenamefont {Son}(2008)}]{Son:2008ye}%
  \BibitemOpen
  \bibfield  {author} {\bibinfo {author} {\bibfnamefont {D.~T.}\ \bibnamefont
  {Son}},\ }\bibfield  {title} {\bibinfo {title} {{Toward an AdS/cold atoms
  correspondence: A Geometric realization of the Schrodinger symmetry}},\
  }\href {https://doi.org/10.1103/PhysRevD.78.046003} {\bibfield  {journal}
  {\bibinfo  {journal} {Phys. Rev. D}\ }\textbf {\bibinfo {volume} {78}},\
  \bibinfo {pages} {046003} (\bibinfo {year} {2008})},\ \Eprint
  {https://arxiv.org/abs/0804.3972} {arXiv:0804.3972 [hep-th]} \BibitemShut
  {NoStop}%
\bibitem [{\citenamefont {Balasubramanian}\ and\ \citenamefont
  {McGreevy}(2008)}]{Balasubramanian:2008dm}%
  \BibitemOpen
  \bibfield  {author} {\bibinfo {author} {\bibfnamefont {K.}~\bibnamefont
  {Balasubramanian}}\ and\ \bibinfo {author} {\bibfnamefont {J.}~\bibnamefont
  {McGreevy}},\ }\bibfield  {title} {\bibinfo {title} {{Gravity duals for
  non-relativistic CFTs}},\ }\href
  {https://doi.org/10.1103/PhysRevLett.101.061601} {\bibfield  {journal}
  {\bibinfo  {journal} {Phys. Rev. Lett.}\ }\textbf {\bibinfo {volume} {101}},\
  \bibinfo {pages} {061601} (\bibinfo {year} {2008})},\ \Eprint
  {https://arxiv.org/abs/0804.4053} {arXiv:0804.4053 [hep-th]} \BibitemShut
  {NoStop}%
\bibitem [{\citenamefont {Kachru}\ \emph {et~al.}(2008)\citenamefont {Kachru},
  \citenamefont {Liu},\ and\ \citenamefont {Mulligan}}]{Kachru:2008yh}%
  \BibitemOpen
  \bibfield  {author} {\bibinfo {author} {\bibfnamefont {S.}~\bibnamefont
  {Kachru}}, \bibinfo {author} {\bibfnamefont {X.}~\bibnamefont {Liu}},\ and\
  \bibinfo {author} {\bibfnamefont {M.}~\bibnamefont {Mulligan}},\ }\bibfield
  {title} {\bibinfo {title} {{Gravity duals of Lifshitz-like fixed points}},\
  }\href {https://doi.org/10.1103/PhysRevD.78.106005} {\bibfield  {journal}
  {\bibinfo  {journal} {Phys. Rev. D}\ }\textbf {\bibinfo {volume} {78}},\
  \bibinfo {pages} {106005} (\bibinfo {year} {2008})},\ \Eprint
  {https://arxiv.org/abs/0808.1725} {arXiv:0808.1725 [hep-th]} \BibitemShut
  {NoStop}%
\bibitem [{\citenamefont {Taylor}(2008)}]{Taylor:2008tg}%
  \BibitemOpen
  \bibfield  {author} {\bibinfo {author} {\bibfnamefont {M.}~\bibnamefont
  {Taylor}},\ }\bibfield  {title} {\bibinfo {title} {{Non-relativistic
  holography}},\ }\href@noop {} {\  (\bibinfo {year} {2008})},\ \Eprint
  {https://arxiv.org/abs/0812.0530} {arXiv:0812.0530 [hep-th]} \BibitemShut
  {NoStop}%
\bibitem [{\citenamefont {Taylor}(2016)}]{Taylor:2015glc}%
  \BibitemOpen
  \bibfield  {author} {\bibinfo {author} {\bibfnamefont {M.}~\bibnamefont
  {Taylor}},\ }\bibfield  {title} {\bibinfo {title} {{Lifshitz holography}},\
  }\href {https://doi.org/10.1088/0264-9381/33/3/033001} {\bibfield  {journal}
  {\bibinfo  {journal} {Class. Quant. Grav.}\ }\textbf {\bibinfo {volume}
  {33}},\ \bibinfo {pages} {033001} (\bibinfo {year} {2016})},\ \Eprint
  {https://arxiv.org/abs/1512.03554} {arXiv:1512.03554 [hep-th]} \BibitemShut
  {NoStop}%
\bibitem [{\citenamefont {Gouteraux}\ and\ \citenamefont
  {Kiritsis}(2011)}]{Gouteraux:2011ce}%
  \BibitemOpen
  \bibfield  {author} {\bibinfo {author} {\bibfnamefont {B.}~\bibnamefont
  {Gouteraux}}\ and\ \bibinfo {author} {\bibfnamefont {E.}~\bibnamefont
  {Kiritsis}},\ }\bibfield  {title} {\bibinfo {title} {{Generalized Holographic
  Quantum Criticality at Finite Density}},\ }\href
  {https://doi.org/10.1007/JHEP12(2011)036} {\bibfield  {journal} {\bibinfo
  {journal} {JHEP}\ }\textbf {\bibinfo {volume} {12}},\ \bibinfo {pages}
  {036}},\ \Eprint {https://arxiv.org/abs/1107.2116} {arXiv:1107.2116 [hep-th]}
  \BibitemShut {NoStop}%
\bibitem [{\citenamefont {Huijse}\ \emph {et~al.}(2012)\citenamefont {Huijse},
  \citenamefont {Sachdev},\ and\ \citenamefont {Swingle}}]{Huijse:2011ef}%
  \BibitemOpen
  \bibfield  {author} {\bibinfo {author} {\bibfnamefont {L.}~\bibnamefont
  {Huijse}}, \bibinfo {author} {\bibfnamefont {S.}~\bibnamefont {Sachdev}},\
  and\ \bibinfo {author} {\bibfnamefont {B.}~\bibnamefont {Swingle}},\
  }\bibfield  {title} {\bibinfo {title} {{Hidden Fermi surfaces in compressible
  states of gauge-gravity duality}},\ }\href
  {https://doi.org/10.1103/PhysRevB.85.035121} {\bibfield  {journal} {\bibinfo
  {journal} {Phys. Rev. B}\ }\textbf {\bibinfo {volume} {85}},\ \bibinfo
  {pages} {035121} (\bibinfo {year} {2012})},\ \Eprint
  {https://arxiv.org/abs/1112.0573} {arXiv:1112.0573 [cond-mat.str-el]}
  \BibitemShut {NoStop}%
\bibitem [{\citenamefont {Bertoldi}\ \emph {et~al.}(2009)\citenamefont
  {Bertoldi}, \citenamefont {Burrington},\ and\ \citenamefont
  {Peet}}]{Bertoldi:2009vn}%
  \BibitemOpen
  \bibfield  {author} {\bibinfo {author} {\bibfnamefont {G.}~\bibnamefont
  {Bertoldi}}, \bibinfo {author} {\bibfnamefont {B.~A.}\ \bibnamefont
  {Burrington}},\ and\ \bibinfo {author} {\bibfnamefont {A.}~\bibnamefont
  {Peet}},\ }\bibfield  {title} {\bibinfo {title} {{Black Holes in
  asymptotically Lifshitz spacetimes with arbitrary critical exponent}},\
  }\href {https://doi.org/10.1103/PhysRevD.80.126003} {\bibfield  {journal}
  {\bibinfo  {journal} {Phys. Rev. D}\ }\textbf {\bibinfo {volume} {80}},\
  \bibinfo {pages} {126003} (\bibinfo {year} {2009})},\ \Eprint
  {https://arxiv.org/abs/0905.3183} {arXiv:0905.3183 [hep-th]} \BibitemShut
  {NoStop}%
\bibitem [{\citenamefont {Mann}(2009)}]{Mann:2009yx}%
  \BibitemOpen
  \bibfield  {author} {\bibinfo {author} {\bibfnamefont {R.~B.}\ \bibnamefont
  {Mann}},\ }\bibfield  {title} {\bibinfo {title} {{Lifshitz Topological Black
  Holes}},\ }\href {https://doi.org/10.1088/1126-6708/2009/06/075} {\bibfield
  {journal} {\bibinfo  {journal} {JHEP}\ }\textbf {\bibinfo {volume} {06}},\
  \bibinfo {pages} {075}},\ \Eprint {https://arxiv.org/abs/0905.1136}
  {arXiv:0905.1136 [hep-th]} \BibitemShut {NoStop}%
\bibitem [{\citenamefont {Balasubramanian}\ and\ \citenamefont
  {McGreevy}(2009)}]{Balasubramanian:2009rx}%
  \BibitemOpen
  \bibfield  {author} {\bibinfo {author} {\bibfnamefont {K.}~\bibnamefont
  {Balasubramanian}}\ and\ \bibinfo {author} {\bibfnamefont {J.}~\bibnamefont
  {McGreevy}},\ }\bibfield  {title} {\bibinfo {title} {{An Analytic Lifshitz
  black hole}},\ }\href {https://doi.org/10.1103/PhysRevD.80.104039} {\bibfield
   {journal} {\bibinfo  {journal} {Phys. Rev. D}\ }\textbf {\bibinfo {volume}
  {80}},\ \bibinfo {pages} {104039} (\bibinfo {year} {2009})},\ \Eprint
  {https://arxiv.org/abs/0909.0263} {arXiv:0909.0263 [hep-th]} \BibitemShut
  {NoStop}%
\bibitem [{\citenamefont {Ayon-Beato}\ \emph {et~al.}(2009)\citenamefont
  {Ayon-Beato}, \citenamefont {Garbarz}, \citenamefont {Giribet},\ and\
  \citenamefont {Hassaine}}]{Ayon-Beato:2009rgu}%
  \BibitemOpen
  \bibfield  {author} {\bibinfo {author} {\bibfnamefont {E.}~\bibnamefont
  {Ayon-Beato}}, \bibinfo {author} {\bibfnamefont {A.}~\bibnamefont {Garbarz}},
  \bibinfo {author} {\bibfnamefont {G.}~\bibnamefont {Giribet}},\ and\ \bibinfo
  {author} {\bibfnamefont {M.}~\bibnamefont {Hassaine}},\ }\bibfield  {title}
  {\bibinfo {title} {{Lifshitz Black Hole in Three Dimensions}},\ }\href
  {https://doi.org/10.1103/PhysRevD.80.104029} {\bibfield  {journal} {\bibinfo
  {journal} {Phys. Rev. D}\ }\textbf {\bibinfo {volume} {80}},\ \bibinfo
  {pages} {104029} (\bibinfo {year} {2009})},\ \Eprint
  {https://arxiv.org/abs/0909.1347} {arXiv:0909.1347 [hep-th]} \BibitemShut
  {NoStop}%
\bibitem [{\citenamefont {Charmousis}\ \emph {et~al.}(2010)\citenamefont
  {Charmousis}, \citenamefont {Gouteraux}, \citenamefont {Kim}, \citenamefont
  {Kiritsis},\ and\ \citenamefont {Meyer}}]{Charmousis:2010zz}%
  \BibitemOpen
  \bibfield  {author} {\bibinfo {author} {\bibfnamefont {C.}~\bibnamefont
  {Charmousis}}, \bibinfo {author} {\bibfnamefont {B.}~\bibnamefont
  {Gouteraux}}, \bibinfo {author} {\bibfnamefont {B.~S.}\ \bibnamefont {Kim}},
  \bibinfo {author} {\bibfnamefont {E.}~\bibnamefont {Kiritsis}},\ and\
  \bibinfo {author} {\bibfnamefont {R.}~\bibnamefont {Meyer}},\ }\bibfield
  {title} {\bibinfo {title} {{Effective Holographic Theories for
  low-temperature condensed matter systems}},\ }\href
  {https://doi.org/10.1007/JHEP11(2010)151} {\bibfield  {journal} {\bibinfo
  {journal} {JHEP}\ }\textbf {\bibinfo {volume} {11}},\ \bibinfo {pages}
  {151}},\ \Eprint {https://arxiv.org/abs/1005.4690} {arXiv:1005.4690 [hep-th]}
  \BibitemShut {NoStop}%
\bibitem [{\citenamefont {Tarrio}\ and\ \citenamefont
  {Vandoren}(2011)}]{Tarrio:2011de}%
  \BibitemOpen
  \bibfield  {author} {\bibinfo {author} {\bibfnamefont {J.}~\bibnamefont
  {Tarrio}}\ and\ \bibinfo {author} {\bibfnamefont {S.}~\bibnamefont
  {Vandoren}},\ }\bibfield  {title} {\bibinfo {title} {{Black holes and black
  branes in Lifshitz spacetimes}},\ }\href
  {https://doi.org/10.1007/JHEP09(2011)017} {\bibfield  {journal} {\bibinfo
  {journal} {JHEP}\ }\textbf {\bibinfo {volume} {09}},\ \bibinfo {pages}
  {017}},\ \Eprint {https://arxiv.org/abs/1105.6335} {arXiv:1105.6335 [hep-th]}
  \BibitemShut {NoStop}%
\bibitem [{\citenamefont {Dong}\ \emph {et~al.}(2012)\citenamefont {Dong},
  \citenamefont {Harrison}, \citenamefont {Kachru}, \citenamefont {Torroba},\
  and\ \citenamefont {Wang}}]{Dong:2012se}%
  \BibitemOpen
  \bibfield  {author} {\bibinfo {author} {\bibfnamefont {X.}~\bibnamefont
  {Dong}}, \bibinfo {author} {\bibfnamefont {S.}~\bibnamefont {Harrison}},
  \bibinfo {author} {\bibfnamefont {S.}~\bibnamefont {Kachru}}, \bibinfo
  {author} {\bibfnamefont {G.}~\bibnamefont {Torroba}},\ and\ \bibinfo {author}
  {\bibfnamefont {H.}~\bibnamefont {Wang}},\ }\bibfield  {title} {\bibinfo
  {title} {{Aspects of holography for theories with hyperscaling violation}},\
  }\href {https://doi.org/10.1007/JHEP06(2012)041} {\bibfield  {journal}
  {\bibinfo  {journal} {JHEP}\ }\textbf {\bibinfo {volume} {06}},\ \bibinfo
  {pages} {041}},\ \Eprint {https://arxiv.org/abs/1201.1905} {arXiv:1201.1905
  [hep-th]} \BibitemShut {NoStop}%
\bibitem [{\citenamefont {Alishahiha}\ \emph {et~al.}(2012)\citenamefont
  {Alishahiha}, \citenamefont {O~Colgain},\ and\ \citenamefont
  {Yavartanoo}}]{Alishahiha:2012qu}%
  \BibitemOpen
  \bibfield  {author} {\bibinfo {author} {\bibfnamefont {M.}~\bibnamefont
  {Alishahiha}}, \bibinfo {author} {\bibfnamefont {E.}~\bibnamefont
  {O~Colgain}},\ and\ \bibinfo {author} {\bibfnamefont {H.}~\bibnamefont
  {Yavartanoo}},\ }\bibfield  {title} {\bibinfo {title} {{Charged Black Branes
  with Hyperscaling Violating Factor}},\ }\href
  {https://doi.org/10.1007/JHEP11(2012)137} {\bibfield  {journal} {\bibinfo
  {journal} {JHEP}\ }\textbf {\bibinfo {volume} {11}},\ \bibinfo {pages}
  {137}},\ \Eprint {https://arxiv.org/abs/1209.3946} {arXiv:1209.3946 [hep-th]}
  \BibitemShut {NoStop}%
\bibitem [{\citenamefont {Gouteraux}\ and\ \citenamefont
  {Kiritsis}(2013)}]{Gouteraux:2012yr}%
  \BibitemOpen
  \bibfield  {author} {\bibinfo {author} {\bibfnamefont {B.}~\bibnamefont
  {Gouteraux}}\ and\ \bibinfo {author} {\bibfnamefont {E.}~\bibnamefont
  {Kiritsis}},\ }\bibfield  {title} {\bibinfo {title} {{Quantum critical lines
  in holographic phases with (un)broken symmetry}},\ }\href
  {https://doi.org/10.1007/JHEP04(2013)053} {\bibfield  {journal} {\bibinfo
  {journal} {JHEP}\ }\textbf {\bibinfo {volume} {04}},\ \bibinfo {pages}
  {053}},\ \Eprint {https://arxiv.org/abs/1212.2625} {arXiv:1212.2625 [hep-th]}
  \BibitemShut {NoStop}%
\bibitem [{\citenamefont {Gath}\ \emph {et~al.}(2013)\citenamefont {Gath},
  \citenamefont {Hartong}, \citenamefont {Monteiro},\ and\ \citenamefont
  {Obers}}]{Gath:2012pg}%
  \BibitemOpen
  \bibfield  {author} {\bibinfo {author} {\bibfnamefont {J.}~\bibnamefont
  {Gath}}, \bibinfo {author} {\bibfnamefont {J.}~\bibnamefont {Hartong}},
  \bibinfo {author} {\bibfnamefont {R.}~\bibnamefont {Monteiro}},\ and\
  \bibinfo {author} {\bibfnamefont {N.~A.}\ \bibnamefont {Obers}},\ }\bibfield
  {title} {\bibinfo {title} {{Holographic Models for Theories with Hyperscaling
  Violation}},\ }\href {https://doi.org/10.1007/JHEP04(2013)159} {\bibfield
  {journal} {\bibinfo  {journal} {JHEP}\ }\textbf {\bibinfo {volume} {04}},\
  \bibinfo {pages} {159}},\ \Eprint {https://arxiv.org/abs/1212.3263}
  {arXiv:1212.3263 [hep-th]} \BibitemShut {NoStop}%
\bibitem [{\citenamefont {Bueno}\ \emph {et~al.}(2014)\citenamefont {Bueno},
  \citenamefont {Chemissany},\ and\ \citenamefont {Shahbazi}}]{Bueno:2012vx}%
  \BibitemOpen
  \bibfield  {author} {\bibinfo {author} {\bibfnamefont {P.}~\bibnamefont
  {Bueno}}, \bibinfo {author} {\bibfnamefont {W.}~\bibnamefont {Chemissany}},\
  and\ \bibinfo {author} {\bibfnamefont {C.~S.}\ \bibnamefont {Shahbazi}},\
  }\bibfield  {title} {\bibinfo {title} {{On $hvLif$-like solutions in gauged
  Supergravity}},\ }\href {https://doi.org/10.1140/epjc/s10052-013-2684-3}
  {\bibfield  {journal} {\bibinfo  {journal} {Eur. Phys. J. C}\ }\textbf
  {\bibinfo {volume} {74}},\ \bibinfo {pages} {2684} (\bibinfo {year}
  {2014})},\ \Eprint {https://arxiv.org/abs/1212.4826} {arXiv:1212.4826
  [hep-th]} \BibitemShut {NoStop}%
\bibitem [{\citenamefont {Pedraza}\ \emph {et~al.}(2019)\citenamefont
  {Pedraza}, \citenamefont {Sybesma},\ and\ \citenamefont
  {Visser}}]{Pedraza:2018eey}%
  \BibitemOpen
  \bibfield  {author} {\bibinfo {author} {\bibfnamefont {J.~F.}\ \bibnamefont
  {Pedraza}}, \bibinfo {author} {\bibfnamefont {W.}~\bibnamefont {Sybesma}},\
  and\ \bibinfo {author} {\bibfnamefont {M.~R.}\ \bibnamefont {Visser}},\
  }\bibfield  {title} {\bibinfo {title} {{Hyperscaling violating black holes
  with spherical and hyperbolic horizons}},\ }\href
  {https://doi.org/10.1088/1361-6382/ab0094} {\bibfield  {journal} {\bibinfo
  {journal} {Class. Quant. Grav.}\ }\textbf {\bibinfo {volume} {36}},\ \bibinfo
  {pages} {054002} (\bibinfo {year} {2019})},\ \Eprint
  {https://arxiv.org/abs/1807.09770} {arXiv:1807.09770 [hep-th]} \BibitemShut
  {NoStop}%
\bibitem [{\citenamefont {Perlmutter}(2012)}]{Perlmutter:2012he}%
  \BibitemOpen
  \bibfield  {author} {\bibinfo {author} {\bibfnamefont {E.}~\bibnamefont
  {Perlmutter}},\ }\bibfield  {title} {\bibinfo {title} {{Hyperscaling
  violation from supergravity}},\ }\href
  {https://doi.org/10.1007/JHEP06(2012)165} {\bibfield  {journal} {\bibinfo
  {journal} {JHEP}\ }\textbf {\bibinfo {volume} {06}},\ \bibinfo {pages}
  {165}},\ \Eprint {https://arxiv.org/abs/1205.0242} {arXiv:1205.0242 [hep-th]}
  \BibitemShut {NoStop}%
\bibitem [{\citenamefont {Brenna}\ \emph {et~al.}(2015)\citenamefont {Brenna},
  \citenamefont {Mann},\ and\ \citenamefont {Park}}]{Brenna:2015pqa}%
  \BibitemOpen
  \bibfield  {author} {\bibinfo {author} {\bibfnamefont {W.~G.}\ \bibnamefont
  {Brenna}}, \bibinfo {author} {\bibfnamefont {R.~B.}\ \bibnamefont {Mann}},\
  and\ \bibinfo {author} {\bibfnamefont {M.}~\bibnamefont {Park}},\ }\bibfield
  {title} {\bibinfo {title} {{Mass and Thermodynamic Volume in Lifshitz
  Spacetimes}},\ }\href {https://doi.org/10.1103/PhysRevD.92.044015} {\bibfield
   {journal} {\bibinfo  {journal} {Phys. Rev. D}\ }\textbf {\bibinfo {volume}
  {92}},\ \bibinfo {pages} {044015} (\bibinfo {year} {2015})},\ \Eprint
  {https://arxiv.org/abs/1505.06331} {arXiv:1505.06331 [hep-th]} \BibitemShut
  {NoStop}%
\bibitem [{\citenamefont {Kiritsis}\ and\ \citenamefont
  {Matsuo}(2017)}]{Kiritsis:2016rcb}%
  \BibitemOpen
  \bibfield  {author} {\bibinfo {author} {\bibfnamefont {E.}~\bibnamefont
  {Kiritsis}}\ and\ \bibinfo {author} {\bibfnamefont {Y.}~\bibnamefont
  {Matsuo}},\ }\bibfield  {title} {\bibinfo {title} {{Hyperscaling-Violating
  Lifshitz hydrodynamics from black-holes: Part II}},\ }\href
  {https://doi.org/10.1007/JHEP03(2017)041} {\bibfield  {journal} {\bibinfo
  {journal} {JHEP}\ }\textbf {\bibinfo {volume} {03}},\ \bibinfo {pages}
  {041}},\ \Eprint {https://arxiv.org/abs/1611.04773} {arXiv:1611.04773
  [hep-th]} \BibitemShut {NoStop}%
\bibitem [{\citenamefont {Kastor}\ \emph {et~al.}(2019)\citenamefont {Kastor},
  \citenamefont {Ray},\ and\ \citenamefont {Traschen}}]{Kastor:2018cqc}%
  \BibitemOpen
  \bibfield  {author} {\bibinfo {author} {\bibfnamefont {D.}~\bibnamefont
  {Kastor}}, \bibinfo {author} {\bibfnamefont {S.}~\bibnamefont {Ray}},\ and\
  \bibinfo {author} {\bibfnamefont {J.}~\bibnamefont {Traschen}},\ }\bibfield
  {title} {\bibinfo {title} {{Black Hole Enthalpy and Scalar Fields}},\ }\href
  {https://doi.org/10.1088/1361-6382/aaf663} {\bibfield  {journal} {\bibinfo
  {journal} {Class. Quant. Grav.}\ }\textbf {\bibinfo {volume} {36}},\ \bibinfo
  {pages} {024002} (\bibinfo {year} {2019})},\ \Eprint
  {https://arxiv.org/abs/1807.09801} {arXiv:1807.09801 [gr-qc]} \BibitemShut
  {NoStop}%
\bibitem [{\citenamefont {Kastor}\ \emph {et~al.}(2009)\citenamefont {Kastor},
  \citenamefont {Ray},\ and\ \citenamefont {Traschen}}]{Kastor:2009wy}%
  \BibitemOpen
  \bibfield  {author} {\bibinfo {author} {\bibfnamefont {D.}~\bibnamefont
  {Kastor}}, \bibinfo {author} {\bibfnamefont {S.}~\bibnamefont {Ray}},\ and\
  \bibinfo {author} {\bibfnamefont {J.}~\bibnamefont {Traschen}},\ }\bibfield
  {title} {\bibinfo {title} {{Enthalpy and the Mechanics of AdS Black Holes}},\
  }\href {https://doi.org/10.1088/0264-9381/26/19/195011} {\bibfield  {journal}
  {\bibinfo  {journal} {Class. Quant. Grav.}\ }\textbf {\bibinfo {volume}
  {26}},\ \bibinfo {pages} {195011} (\bibinfo {year} {2009})},\ \Eprint
  {https://arxiv.org/abs/0904.2765} {arXiv:0904.2765 [hep-th]} \BibitemShut
  {NoStop}%
\bibitem [{\citenamefont {Kubiznak}\ \emph {et~al.}(2017)\citenamefont
  {Kubiznak}, \citenamefont {Mann},\ and\ \citenamefont
  {Teo}}]{Kubiznak:2016qmn}%
  \BibitemOpen
  \bibfield  {author} {\bibinfo {author} {\bibfnamefont {D.}~\bibnamefont
  {Kubiznak}}, \bibinfo {author} {\bibfnamefont {R.~B.}\ \bibnamefont {Mann}},\
  and\ \bibinfo {author} {\bibfnamefont {M.}~\bibnamefont {Teo}},\ }\bibfield
  {title} {\bibinfo {title} {Black hole chemistry: thermodynamics with
  lambda},\ }\href {http://stacks.iop.org/0264-9381/34/i=6/a=063001} {\bibfield
   {journal} {\bibinfo  {journal} {Classical and Quantum Gravity}\ }\textbf
  {\bibinfo {volume} {34}},\ \bibinfo {pages} {063001} (\bibinfo {year}
  {2017})}\BibitemShut {NoStop}%
\bibitem [{\citenamefont {Romero-Figueroa}\ and\ \citenamefont
  {Quevedo}(2024)}]{Romero-Figueroa:2024sac}%
  \BibitemOpen
  \bibfield  {author} {\bibinfo {author} {\bibfnamefont {C.~E.}\ \bibnamefont
  {Romero-Figueroa}}\ and\ \bibinfo {author} {\bibfnamefont {H.}~\bibnamefont
  {Quevedo}},\ }\bibfield  {title} {\bibinfo {title} {{Extended thermodynamics
  and critical behavior of generalized dilatonic Lifshitz black holes}},\
  }\href@noop {} {\  (\bibinfo {year} {2024})},\ \Eprint
  {https://arxiv.org/abs/2406.18223} {arXiv:2406.18223 [gr-qc]} \BibitemShut
  {NoStop}%
\bibitem [{\citenamefont {Mancilla}(2024)}]{Mancilla:2024spp}%
  \BibitemOpen
  \bibfield  {author} {\bibinfo {author} {\bibfnamefont {R.}~\bibnamefont
  {Mancilla}},\ }\bibfield  {title} {\bibinfo {title} {{Generalized Euler
  Equation from Effective Action: Implications for the Smarr Formula in AdS
  Black Holes}},\ }\href@noop {} {\  (\bibinfo {year} {2024})},\ \Eprint
  {https://arxiv.org/abs/2410.06605} {arXiv:2410.06605 [hep-th]} \BibitemShut
  {NoStop}%
\bibitem [{\citenamefont {Ross}\ and\ \citenamefont
  {Saremi}(2009)}]{Ross:2009ar}%
  \BibitemOpen
  \bibfield  {author} {\bibinfo {author} {\bibfnamefont {S.~F.}\ \bibnamefont
  {Ross}}\ and\ \bibinfo {author} {\bibfnamefont {O.}~\bibnamefont {Saremi}},\
  }\bibfield  {title} {\bibinfo {title} {{Holographic stress tensor for
  non-relativistic theories}},\ }\href
  {https://doi.org/10.1088/1126-6708/2009/09/009} {\bibfield  {journal}
  {\bibinfo  {journal} {JHEP}\ }\textbf {\bibinfo {volume} {09}},\ \bibinfo
  {pages} {009}},\ \Eprint {https://arxiv.org/abs/0907.1846} {arXiv:0907.1846
  [hep-th]} \BibitemShut {NoStop}%
\bibitem [{\citenamefont {Mann}\ and\ \citenamefont
  {McNees}(2011)}]{Mann:2011hg}%
  \BibitemOpen
  \bibfield  {author} {\bibinfo {author} {\bibfnamefont {R.~B.}\ \bibnamefont
  {Mann}}\ and\ \bibinfo {author} {\bibfnamefont {R.}~\bibnamefont {McNees}},\
  }\bibfield  {title} {\bibinfo {title} {{Holographic Renormalization for
  Asymptotically Lifshitz Spacetimes}},\ }\href
  {https://doi.org/10.1007/JHEP10(2011)129} {\bibfield  {journal} {\bibinfo
  {journal} {JHEP}\ }\textbf {\bibinfo {volume} {10}},\ \bibinfo {pages}
  {129}},\ \Eprint {https://arxiv.org/abs/1107.5792} {arXiv:1107.5792 [hep-th]}
  \BibitemShut {NoStop}%
\bibitem [{\citenamefont {Chemissany}\ and\ \citenamefont
  {Papadimitriou}(2015)}]{Chemissany:2014xsa}%
  \BibitemOpen
  \bibfield  {author} {\bibinfo {author} {\bibfnamefont {W.}~\bibnamefont
  {Chemissany}}\ and\ \bibinfo {author} {\bibfnamefont {I.}~\bibnamefont
  {Papadimitriou}},\ }\bibfield  {title} {\bibinfo {title} {{Lifshitz
  holography: The whole shebang}},\ }\href
  {https://doi.org/10.1007/JHEP01(2015)052} {\bibfield  {journal} {\bibinfo
  {journal} {JHEP}\ }\textbf {\bibinfo {volume} {01}},\ \bibinfo {pages}
  {052}},\ \Eprint {https://arxiv.org/abs/1408.0795} {arXiv:1408.0795 [hep-th]}
  \BibitemShut {NoStop}%
\bibitem [{\citenamefont {Cardy}(1986)}]{Cardy:1986ie}%
  \BibitemOpen
  \bibfield  {author} {\bibinfo {author} {\bibfnamefont {J.~L.}\ \bibnamefont
  {Cardy}},\ }\bibfield  {title} {\bibinfo {title} {{Operator Content of
  Two-Dimensional Conformally Invariant Theories}},\ }\href
  {https://doi.org/10.1016/0550-3213(86)90552-3} {\bibfield  {journal}
  {\bibinfo  {journal} {Nucl. Phys. B}\ }\textbf {\bibinfo {volume} {270}},\
  \bibinfo {pages} {186} (\bibinfo {year} {1986})}\BibitemShut {NoStop}%
\bibitem [{\citenamefont {Gonzalez}\ \emph {et~al.}(2011)\citenamefont
  {Gonzalez}, \citenamefont {Tempo},\ and\ \citenamefont
  {Troncoso}}]{Gonzalez:2011nz}%
  \BibitemOpen
  \bibfield  {author} {\bibinfo {author} {\bibfnamefont {H.~A.}\ \bibnamefont
  {Gonzalez}}, \bibinfo {author} {\bibfnamefont {D.}~\bibnamefont {Tempo}},\
  and\ \bibinfo {author} {\bibfnamefont {R.}~\bibnamefont {Troncoso}},\
  }\bibfield  {title} {\bibinfo {title} {{Field theories with anisotropic
  scaling in 2D, solitons and the microscopic entropy of asymptotically
  Lifshitz black holes}},\ }\href {https://doi.org/10.1007/JHEP11(2011)066}
  {\bibfield  {journal} {\bibinfo  {journal} {JHEP}\ }\textbf {\bibinfo
  {volume} {11}},\ \bibinfo {pages} {066}},\ \Eprint
  {https://arxiv.org/abs/1107.3647} {arXiv:1107.3647 [hep-th]} \BibitemShut
  {NoStop}%
\bibitem [{\citenamefont {Shaghoulian}(2015)}]{Shaghoulian:2015dwa}%
  \BibitemOpen
  \bibfield  {author} {\bibinfo {author} {\bibfnamefont {E.}~\bibnamefont
  {Shaghoulian}},\ }\bibfield  {title} {\bibinfo {title} {{A Cardy formula for
  holographic hyperscaling-violating theories}},\ }\href
  {https://doi.org/10.1007/JHEP11(2015)081} {\bibfield  {journal} {\bibinfo
  {journal} {JHEP}\ }\textbf {\bibinfo {volume} {11}},\ \bibinfo {pages}
  {081}},\ \Eprint {https://arxiv.org/abs/1504.02094} {arXiv:1504.02094
  [hep-th]} \BibitemShut {NoStop}%
\bibitem [{\citenamefont {Bravo~Gaete}\ \emph {et~al.}(2017)\citenamefont
  {Bravo~Gaete}, \citenamefont {Guajardo},\ and\ \citenamefont
  {Hassaine}}]{BravoGaete:2017dso}%
  \BibitemOpen
  \bibfield  {author} {\bibinfo {author} {\bibfnamefont {M.}~\bibnamefont
  {Bravo~Gaete}}, \bibinfo {author} {\bibfnamefont {L.}~\bibnamefont
  {Guajardo}},\ and\ \bibinfo {author} {\bibfnamefont {M.}~\bibnamefont
  {Hassaine}},\ }\bibfield  {title} {\bibinfo {title} {{A Cardy-like formula
  for rotating black holes with planar horizon}},\ }\href
  {https://doi.org/10.1007/JHEP04(2017)092} {\bibfield  {journal} {\bibinfo
  {journal} {JHEP}\ }\textbf {\bibinfo {volume} {04}},\ \bibinfo {pages}
  {092}},\ \Eprint {https://arxiv.org/abs/1702.02416} {arXiv:1702.02416
  [hep-th]} \BibitemShut {NoStop}%
\end{thebibliography}%

\end{document}